\documentclass[9pt,twocolumn,twoside]{pnas-new-mod}

\usepackage{siunitx}
\usepackage{hyperref}
\usepackage{natbib}

\usepackage[defaultcolor=red,final]{changes} 

\templatetype{pnasresearcharticle} 

\title{\vspace*{-0.15in} Modular DNA origami-based electrochemical detection of DNA and proteins}

\author[a,1,2]{Byoung-jin Jeon}
\author[a,1]{Matteo M. Guareschi}
\author[b]{Jaimie M. Stewart}
\author[c]{Emily Wu}
\author[c]{Ashwin Gopinath}
\author[d]{Netzahualc\'{o}yotl Arroyo-Curr\'{a}s}
\author[e]{Philippe Dauphin-Ducharme}
\author[f]{Kevin W. Plaxco}
\author[g,2]{Philip S. Lukeman}
\author[a,2]{Paul W. K. Rothemund}

\affil[a]{Department of Bioengineering, California Institute of Technology}
\affil[b]{Department of Bioengineering, University of California, Los Angeles}
\affil[c]{Department of Mechanical Engineering, Massachusetts Institute of Technology}
\affil[d]{Department of Pharmacology and Molecular Sciences, Johns Hopkins University School of Medicine}
\affil[e]{D\'{e}partement de chimie, Universit\'{e} de Sherbrooke}
\affil[f]{Department of Chemistry and Biochemistry, University of California Santa Barbara}
\affil[g]{Department of Chemistry, St. John's University }

\leadauthor{Jeon} 

\authorcontributions{Author contributions: A.G. developed original concepts for conformation-switching origami sensors in P.W.K.R.'s lab. P.S.L. developed original concepts for conformation-switching origami sensors in collaboration with K.W.P.'s lab. Discussions between authors B.J., M.M.G., J.M.S., K.W.P, P.D.-D., N.A.-C., P.S.L. and P.W.K.R. resulted in the particular conformation-switching molecular designs and electrochemical experimental design described in this work. Experiments were conducted by B.J., M.M.G., J.M.S. and E.W. Analysis of experiments and first-draft paper-writing were conducted by B.J. and M.M.G. All authors were involved in subsequent editing of the manuscript.}
\authordeclaration{Competing interests: A patent related to this work with inventors A.G. and P.W.K.R. was issued as U.S. patent 11,391,734 to California Institute of Technology and licensed to Somalogic, Inc.}
\correspondingauthor{\vspace*{-0.08in}\textsuperscript{1}B.J. and M.M.G. contributed equally to this work.

\textsuperscript{2}To whom correspondence should be addressed. \\ E-mail: bjeon83@gmail.com, lukemanp@stjohns.edu, pwkr@dna.caltech.edu
}

\keywords{\vspace*{-0.1in}DNA origami $|$ biosensor $|$ square-wave voltammetry $|$ modular sensor } 

\begin{abstract}
\textcolor{blue}{Significance --- Aptamer-based, conformation-switching electrochemical biosensors enable real-time analyte sensing in the living body. However, developing sensors for new analytes is limited by the difficulty of optimizing electrochemical signal strength and gain, without compromising the binding properties of the aptamer. Optimization of new sensors often requires synthesis of many expensive, multiply-modified oligonucleotides. In contrast, we describe a reagentless DNA origami-based platform whose binding and signaling properties are essentially independent. This independence enables signal optimization to be based on analyte size; sensors are customized using a library of inexpensive and unmodified DNA linkers. This library is portable between different sensor designs and analytes. Our work thus provides a ``toolkit'' for constructing electrochemical biosensors, and a general approach for modular biosensors beyond electrochemical detection. \\}

\vspace*{-0.1in}
Abstract --- The diversity and heterogeneity of biomarkers has made the development of general methods for single-step quantification of analytes difficult. For individual biomarkers, electrochemical methods that detect a conformational change in an affinity binder upon analyte binding have shown promise. However, because the conformational change must operate within a nanometer-scale working distance, an entirely new sensor, with a unique conformational change, must be developed for each analyte. Here, we demonstrate a modular electrochemical biosensor, built from DNA origami, which is easily adapted to diverse molecules by merely replacing its analyte binding domains. Instead of relying on a unique nanometer-scale movement of a single redox reporter, all sensor variants rely on the same 100-nanometer scale conformational change, which brings dozens of reporters close enough to a gold electrode surface that a signal can be measured via square wave voltammetry, a standard electrochemical technique. To validate our sensor's mechanism, we used single-stranded DNA as an analyte, and optimized the number of redox reporters and various linker lengths. Adaptation of the sensor to streptavidin and PDGF-BB analytes was achieved by simply adding biotin or anti-PDGF aptamers to appropriate DNA linkers. Geometrically-optimized streptavidin sensors exhibited signal gain and limit of detection markedly better than comparable reagentless electrochemical sensors. After use, the same sensors could be regenerated under mild conditions: performance was largely maintained over four cycles of DNA strand displacement and rehybridization. By leveraging the modularity of DNA nanostructures, our work provides a straightforward route to the single-step quantification of arbitrary nucleic acids and proteins.

\end{abstract}

\dates{This manuscript was compiled on \today}

\begin{document}

\maketitle
\thispagestyle{firststyle}
\ifthenelse{\boolean{shortarticle}}{\ifthenelse{\boolean{singlecolumn}}{\abscontentformatted}{\abscontent}}{}

\vspace*{-0.15in}

\dropcap{D}NA nanotechnology\cite{JonesSeemanMirkin2015,SeemanSleiman2017} enables the design and construction of artificial structures with nanometer precision via sequence-specific assembly of DNA oligonucleotides. Such structures ({\em e.g.} branched DNAs and tetrahedra, \citenum{IdiliBranchedYSensor2022, pei_regenerable_2011,li_design_2017,TetrahedronReview}) are often used as scaffolds for biosensors, wherein DNA-coupled binding components, such as aptamers, antibodies or small-molecule ligands produce a signal upon introduction of a cognate analyte molecule. DNA origami\cite{rothemund_folding_2006, DouglasShih3DOrigami2009}, provides an architecture for making 100-nanometer scale structures of arbitrary geometry, with hundreds of attachment sites for active components such as binders or reporters. Thus DNA origami is of particular interest for biosensors\cite{WangOrigamiEnabledBiosensors2020,YonggangOrigamiSensorsTherapeutics2021,DingNanodevices2022} because it allows the synthesis of reconfigurable structures that undergo large  conformational changes (tens or hundreds of nanometers,  \citenum{kuzuya_nanomechanical_2011, kuzuya_nanomechanical_2014, EndoWillner2015ZnSensorReconfiguration, EndoSugiyamaMao2014, CastroEntropicSensor2017, 
Schickinger2018, WongShihNanocalipers2021, WongNILSA2017, WongHalvorsen2021NanoswitchBarcodes, andersen_self-assembly_2009, TangHeddleTannerOrigamiAptamerBox2018, AndersenFluorescentOrigamiBeacon2018, KuzykLiedlLiu2014, 
KuzykLiedlRNAChiralPlasmonic2018, NaLiuAptamer2018, KuzykOrigamiChiralSensing2018}), rearranging large numbers  of reporters (tens or hundreds, \citenum{AndersenFluorescentOrigamiBeacon2018}) that amplify molecular binding events to the point that they can be reliably detected. 

Reconfigurable origami sensor platforms can be divided into broad classes based on their readout mechanism, with each mechanism having advantages and disadvantages.
Atomic force microscope (AFM)-based platforms, for example, such as ``DNA forceps''
\cite{kuzuya_nanomechanical_2011, kuzuya_nanomechanical_2014} or ``picture frames'' \cite{EndoWillner2015ZnSensorReconfiguration} enable the direct visualization of single molecule reconfiguration as a function of protein binding\cite{kuzuya_nanomechanical_2011}, pH\cite{kuzuya_nanomechanical_2014}, or ionic conditions\cite{EndoWillner2015ZnSensorReconfiguration} but they require expensive AFMs and time-consuming measurements. Single-molecule fluorescence, optical tweezers, and magnetic tweezers-based platforms\cite{EndoSugiyamaMao2014, CastroEntropicSensor2017, Schickinger2018, WongShihNanocalipers2021} exhibit extraordinary temporal resolution for the reconfigurations of single molecules, and enable the dissection of sensor states that could otherwise not be resolved. But, as for AFM, the required equipment is expensive and the measurements time-consuming.
Gel electrophoresis-based platforms\cite{WongNILSA2017} for protein or nucleic acid detection are relatively inexpensive, and can be multiplexed \cite{WongHalvorsen2021NanoswitchBarcodes}. However, these measurements are slow, at best taking tens of minutes.
Fluorescence resonance energy transfer (FRET) platforms have provided simple bulk measurements of DNA\cite{andersen_self-assembly_2009} and protein\cite{TangHeddleTannerOrigamiAptamerBox2018} down to the 100~nM range; achieving 100~pM has required expensive single-molecule FRET microscopy\cite{AndersenFluorescentOrigamiBeacon2018}.
 Chiral plasmonic platforms have enabled simple bulk sensing of nucleic acids\cite{KuzykLiedlLiu2014, 
KuzykLiedlRNAChiralPlasmonic2018} and small molecules\cite{NaLiuAptamer2018, KuzykOrigamiChiralSensing2018}  achieving 100~pM sensitivity for RNA, but spectrometers capable of measuring circular dichroism are relatively expensive. Layered on top of these concerns, few of these approaches have been shown to work in complex biological sample matrices, and none of the above platforms appear easily extendable to {\em in vivo} measurements. 

\clearpage

In contrast to these techniques, an electrochemical platform based on the conformation-switching of single-stranded DNAs\cite{Netz2020Beaker2Body} enables real-time measurements of analytes using inexpensive potentiostats \cite{SmartphonePotentiostatSoleymani2022}.  By using miniaturized electrode implants, analytes can be quantified in challenging environments such as living animals\cite{NetzPlaxcoAwake2017, PlaxcoDruggedRats2023}. This platform uses a gold electrode, protected with an alkanethiol monolayer---the electrode is further functionalized with a target-binding oligonucleotide that displays a gold-binding thiol on one end of the oligonucleotide, and a redox-reporter (most commonly methylene blue, ``MB'') distal to the thiol. 
Upon interacting with its target, for example during DNA hybridization (``E-DNA sensors'', \citenum{Fan2003a,Xiao2006, Barton2018reviewDNAsensing}) or aptamer-molecule binding (``E-AB sensors'', \citenum{xiao_label-free_2005, Hu2012a,NetzReview2020}), the oligonucleotide undergoes changes that affect the rate of electron transfer between the reporter and the electrode surface. When the electrode is brought to the reduction/oxidation potential of the reporter, current is thus measured as a function of binding state of the oligonucleotide.

In the most common model of E-AB sensor operation, analyte binding-induced reconfiguration of the aptamer structure changes the collision frequency of the redox reporter with the electrode surface, changing the electron transfer rate and thus measured current\cite{PlaxcoFundamentalAptamerSensorMechanisms2009}.  Analyte binding in E-AB sensors is transduced into a signal by a unique and idiosyncratic binding-induced nanometer-scale movement; this conformational change {\em differs between different aptamers} ({\em e.g.} see differences between thrombin and IgE aptamers in ref.~\citenum{PlaxcoFundamentalAptamerSensorMechanisms2009}) and it is evident that binding and signal transduction depend on each other in complex ways. Thus, making  sensors for new analytes with both good signal gain (relative change in signal upon target saturation) and appropriate affinity and selectivity often requires optimization, with multiple rounds of semiempirical redesign and resynthesis of expensive modified oligonucleotides \cite{PlaxcoReingineeringAptamerSensors2010}. Spectroscopy-guided approaches to aptamer probe redesign \cite{PlaxcoStojanovicReengineeringAptamers2023} shorten, but do not eliminate this onerous process. Special ``capture SELEX'' techniques that select for conformational change of the aptamer\cite{CaptureSelex2012, Stojanovic2016CaptureSelex}  rather than simple binding, can increase the likelihood of obtaining suitable switching aptamers but cannot guarantee success.

The question arises: might the advantages of reconfigurable DNA origami and E-DNA/E-AB systems be combined? Several recent publications explore DNA origami in the context of electrochemical sensing. For example, the capacity of origami to display multiple binders has been used for an electrode-bound miRNA sensor, wherein origami acts as a substrate that presents multiple miRNA binding sites\cite{han_facile_2019}. Likewise, a pH sensor has been reported that employs a reconfigurable DNA origami ``zipper'' bound to a gold electrode\cite{williamsonProbingConformationalStates2021}, and electrode-bound origami were used to probe the spatial dependence of redox-active enzyme activity\cite{ge_constructing_2019-1}. Free DNA origami rectangles have been used to amplify signal from DNA analytes\cite{ElectrochemOrigami2023}; the detection of free origami rectangles themselves represent an elegant demonstration of nanoimpact electrochemistry\cite{PensaSimmel2022}. Unlike E-AB/E-DNA systems, however, these origami-based systems are not reagentless; they all require the addition of redox mediators to the analyte solution to generate electrochemical signals, which limits their use {\em in vivo} or in other environments where these mediators are not available. And aside from the pH-responsive zipper\cite{williamsonProbingConformationalStates2021}, none of the above systems use large-scale conformational reconfigurations for signal transduction. 

One reconfigurable origami-based electrochemical sensor that uses MB reporters instead of added redox mediators, has been reported \cite{arroyo-curras_electrochemical_2020}. Intended to detect 100~nm scale analytes such as viruses, this system requires significant modification for analytes of different sizes. Furthermore, as configured, it is a ``signal-off'' device, for which analyte binding results in a lower current; this limits both the signal gain (at best, -100\%; see formula for gain below) and confidence that signal change is not caused by sensor degradation.

\begin{figure*}
    \centering
    \includegraphics[width=0.8\textwidth]{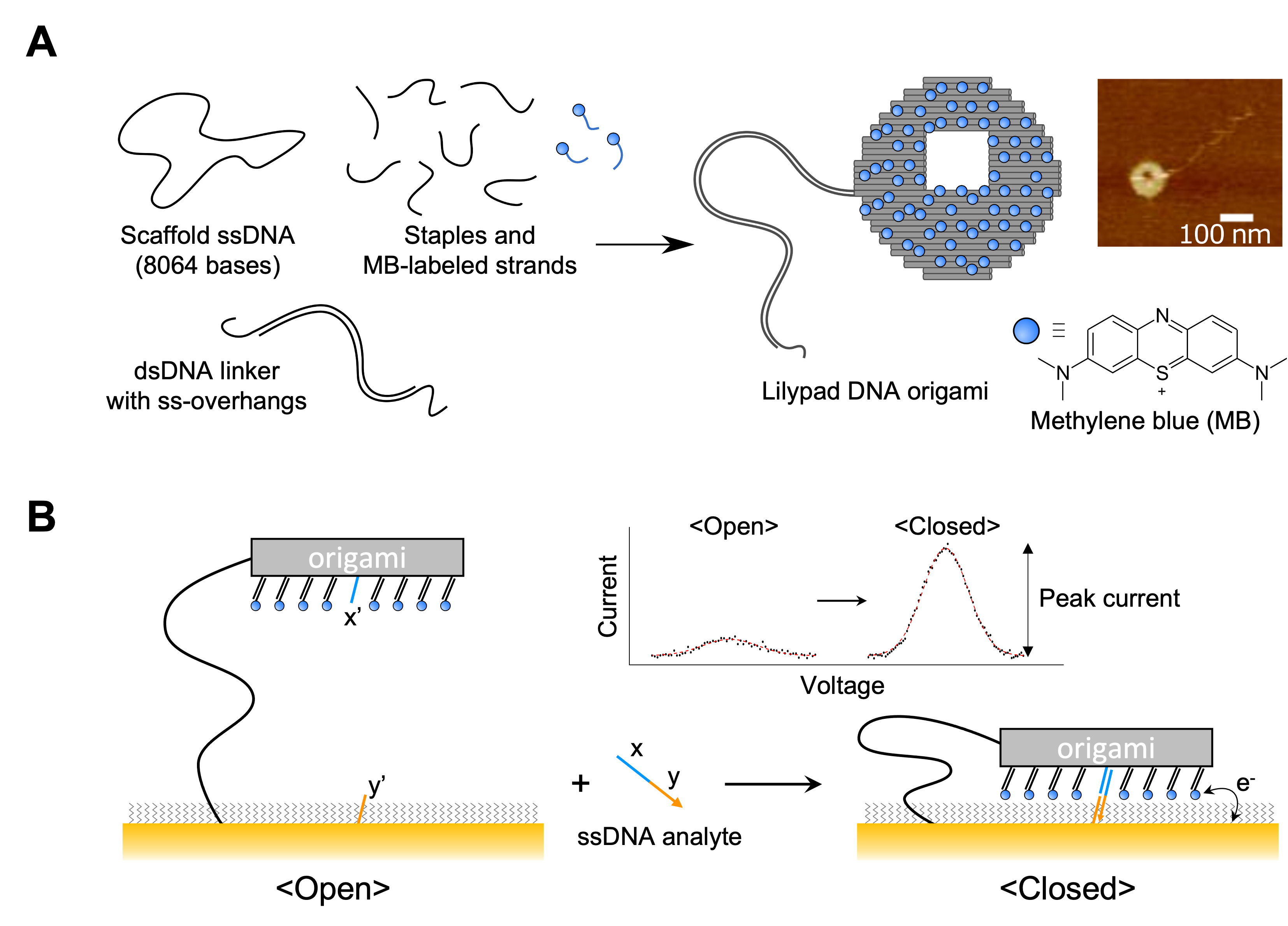}
    \caption{ \textbf{Lily pad sensors can be used for the electrochemical detection of biological analytes, here a DNA single strand.} (A) A flat, disk-shaped DNA origami with a square hole \cite{gopinathAbsoluteArbitraryOrientation2021} carrying a long dsDNA linker is assembled from a mixture of long ssDNA scaffolds, ssDNA staples, 70~staples with 5$^\prime$-extensions for MB reporters and for analyte binding, 20~nt MB-modified DNA oligos, and long dsDNA linkers with ssDNA overhangs at both ends. Inset shows AFM of a lily pad. (B) Two types of thiol-modified ssDNA are immobilized on a template-stripped gold surface: one for tethering the lily pad origami via hybridization with one of the linker overhang sequences, and the other for analyte binding to the gold surface. Closing events occur when a DNA analyte binds both to the binding site on the origami and the binding site on the surface, enabling electron transfer between MB reporters and the gold electrode. Inset shows an exemplary square wave voltammogram of a sensor in the open state (left), and a voltammogram in the closed state after DNA analyte addition (right). The peak current is calculated as the heights of the peaks relative to the underlying baseline.}
    \label{fig:1}
\end{figure*}

In this work, we introduce a reconfigurable DNA origami sensor whose modular architecture overcomes the redesign and optimization challenges posed by E-DNA and E-AB sensors. By combining a flat, two-dimensional DNA origami with a double-stranded DNA (dsDNA) linker, we create a ``lily pad'' structure wherein origami, decorated with numerous MB redox reporters are tethered to an ultra-flat gold electrode via a long and flexible dsDNA linker (Fig.~1). This open conformation is converted to a closed conformation by the presence of analyte biomolecules when they bind and form a bridge between a pair of probes on the origami and on the electrode surface. The resulting conformational change is detected via square-wave voltammetry (SWV, Fig.~1B, inset), in which the current, which is a function of the rate of electron transfer between MB molecules and the electrode surface, increases with proximity.

The lily pad architecture described above is reminiscent of a classical sandwich assay, in which a pair of affinity binders is used to localize a signaling system to a surface if and only if the target analyte attaches to both binders: the first binder, or ``capture'' reagent, is attached to a surface; addition of sample results in target analytes being bound to capture reagents via a first epitope; further addition of a labeled ``detect'' reagent that binds the analyte at a second orthogonal epitope provides a mechanism to generate signal. Simple binders, rather than the structure-switching binders of E-AB systems, are all that is required; industry has generated many thousands of capture-detect antibody pairs for commercial assays. Indeed, antibody pairs have been incorporated into a number of electrochemical assays---yet these systems are not reagentless, typically requiring both (1) a step involving physical addition of the detect reagent and (2) incorporation of an amplification system (typically attached to the detect reagent) such as an enzyme \cite{electrochemicalSandwich2021review}, or hybridization chain reaction \cite{HCRelectrochemicalSandwich2018}. 

Given their success elsewhere in biosensing, it is interesting to ask why pairs of sandwich antibodies have not proven amenable to reagentless electrochemical sensing. Reagentless sensors require (1) that all of their components, from the binders to any detection system to be built into a single, intramolecular device, which reconfigures upon analyte binding and (2) that the reconfiguration be able to work with the detection system to create a signal of sufficient magnitude. The first requirement might in principle be solved by a simple linker between the binders, but the second requirement turns out to be a more fundamental problem. In typical sandwich immunoassays, the stack formed by the capture reagent, analyte, and detect reagent is at least several nanometers in height, {\em e.g.} 10~nm in the case of bovine serum albumin with two antibodies\cite{MurphyAntibodyStackSize1988}. To render the device reagentless, some component has to be directly labeled with a redox reporter; labeling either antibody would result in a geometry for which electron transfer to the surface is too slow for sensitive detection (for E-AB sensors, electron transfer rates drop off significantly a few nm from the electrode surface \cite{ElectronTransferChainDymics2018}). \added{Using smaller binders in sandwich assays, {\em e.g.} nanobodies\cite{Harmsen2007}, could help increase the range of accessible analytes, but the use of smaller binders in this context merely shifts the limiting factor to analyte size.} This length scale problem has limited development of reagentless electrochemical sandwich immunoassays. A cleverly-designed system\cite{Sandwich2012}  utilizes a combination of DNA linkers with an antibody pair, in a geometry that positions a MB reporter at the very base of the sensor in the analyte-bound state. This results in  an antibody sandwich acting as a highly sensitive, amplification-free, electrochemical sensor whose performance does not depend strongly on the size of the binders or analyte. However, that sensor still requires addition of DNA-labeled antibodies in solution and thus does not operate in a reagentless mode. 

Here, our lily pad design overcomes the problem of large binder-analyte sandwiches through its use of numerous MB reporters which project down from the origami in a flexible ``curtain'' around the sandwich that extends at least 5~nm towards the surface. Like E-DNA/E-AB systems, our lily pad incorporates all components in a single, intramolecular construct rendering it reagentless. Unlike the E-DNA/E-AB systems the lily pad is more easily adapted to other analytes---by simply exchanging unmodified strands for linkers and curtain length we demonstrate the detection of DNA and two different proteins.

\section*{Results}

\subsection*{Lily pad DNA origami device}

\added{Figure~1 shows the design of our lily pad device. The device consists of a 100~nm diameter disk-shaped DNA single layer origamidesigned on a square lattice, starting from a previously published model\cite{gopinathAbsoluteArbitraryOrientation2021}. During folding (see methods), the origami is attached to a long dsDNA linker (Fig.~1A) which, depending on experiment, ranges from 300~to 3000~base pairs (bp) in length. The distal end of the linker is attached to a gold electrode through DNA hybridization: the distal end of the linker displays a single-stranded DNA (ssDNA) overhang, and binds to a complementary ssDNA that is immobilized on the gold electrode.} E-DNA/E-AB sensors often use roughened gold electrodes to maximize the surface area for their small ssDNA sensing molecules\cite{arroyo-curras_high_2017,mahshidMechanisticControlGrowth2016},  despite the resultant structural heterogeneity of such electrodes. Here, in contrast, we use ultraflat, template-stripped gold chips\cite{hegnerUltralargeAtomicallyFlat1993} to minimize the distance between the entire bottom face of the origami disk and the electrode surface upon analyte binding. Our lily pad sensors use 70~MB molecules as reporters to transfer electrons to the electrode; they are attached to one side of each origami via 20~bp-long ssDNAs hybridized to overhangs on 70~modified staple strands (Fig.~1B)---this forms a curtain of MB-modified DNA strands hanging from the origami. Detection of a ssDNA analyte sequence \textit{xy} is achieved through binding to two DNA sequences, \textit{x}$^\prime$ and \textit{y}$^\prime$,  which are complementary to subsequences \textit{x} and \textit{y} of \textit{xy}, respectively. Sequence \textit{x}$^\prime$ is positioned on the origami as an extension to a central staple, and a thiol-modified version of \textit{y}$^\prime$ is immobilized on the gold electrode surface via an \mbox{Au-S} bond. When both \textit{x} and \textit{y} in a DNA analyte bind to \textit{x}$^\prime$ and \textit{y}$^\prime$ on the lily pad and the underlying electrode, the conformation of the lily pad changes from the ``open'' to the ``closed'' state. Each closing event brings the 70~MB curtain into proximity with the gold surface, facilitating electron transfer between MB and electrode and increasing voltammetric peak currents (Fig.~1B, inset). 

The distance between MB molecules and the electrodes, which determines the electron transfer rate, is set by the particular probe-analyte binding geometry \cite{dauphin-ducharme_high-precision_2019-3}. We note that in the open, unbound state, the relative average position of the MB redox reporters with respect to the surface is a function of the flexibilities of the dsDNA linker, origami\cite{niDirectVisualizationFloppy2022,leeRapidComputationalAnalysis2021} and MB attachment; therefore we would expect background signal even in the absence of analyte (see below). Background signal is observed with E-DNA and E-AB sensors for similar reasons \cite{Xiao2006,pandeyIntegratingProgrammableDNAzymes2021}, and it is observed here (Fig.~1B, left side of inset). 

For the closed, bound state, the same factors should affect electron transfer rate, with the exception that the distance between the lily-pad's reporters and the surface should be independent of the dsDNA linker length. We estimate the reporter-surface distance for the closed state by noting that the length of the bound DNA analyte is 28~bp ($\sim$9.8 nm);  as the tail length is 20~bp ($\sim$6.8 nm)  for the reporter strands, this difference leaves the redox reporters nominally $\sim$3.0~nm away from the surface on average (Fig.~2A, blue inset). Were the MB rigidly held at this position, thirty times the electron-transfer decay distance \cite{Dauphin-Ducharme2017a}, electron transfer rates would be unmeasurably slow; thus we suspect that bending fluctuations of the origami cause MB to visit distances less than 3~nm from the surface, enabling electron transfer from the MB curtain to be observed.

In principle, more MB units per lily pad should generate a greater signal upon analyte binding. However, saturation MB modification (1~MB per staple) leads to aggregation of origami during annealing, potentially due to the DNA intercalation of MB, electrostatic DNA backbone-MB interactions, or concentration-dependent dimerization of MB \cite{norden_structure_1982-1,bradley_electric_1972}. Optimizing lily pad MB density while minimizing aggregation, we found that the use of 70~MB-modified extensions results in lily pads that run as monomers in agarose gels, while higher numbers of MB reporters lead to poorly formed structures that aggregate and remain stuck in gel wells (for optimization see Table~S1 and Fig.~S1).

\begin{figure*}

    \centering
    \includegraphics[width=0.88\textwidth]{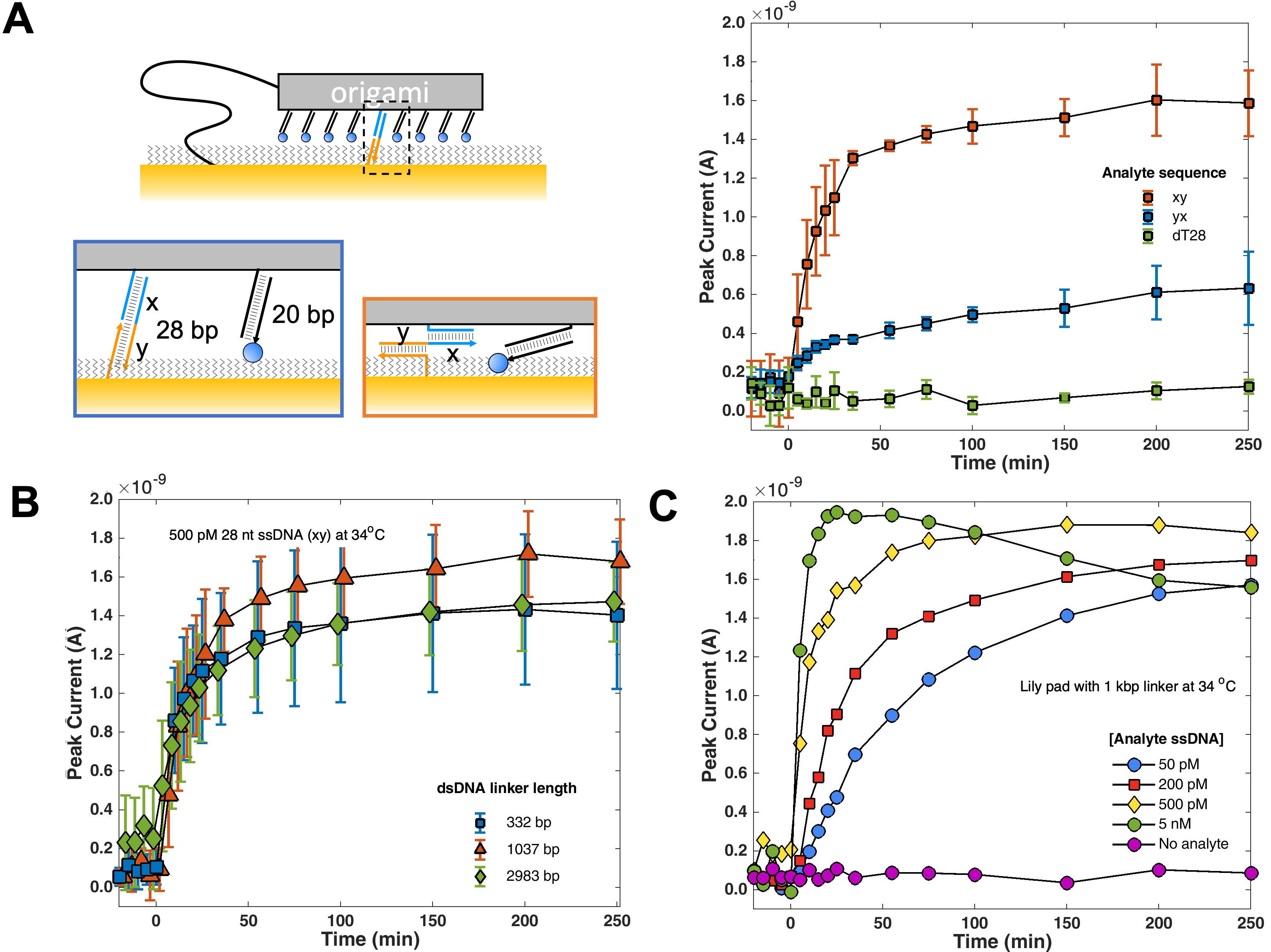}
    \caption{ \textbf{Lily pad closing events can be monitored via voltammetry in real-time.} \added{(A)  500~pM of each of three different ssDNA analytes, with 28-bp sequences \textit{xy} (blue), \textit{yx} (orange), and \textit{dT} (green) were added at time $t = 0$ at 34$^\circ$C. The difference in kinetics  for \textit{xy} and \textit{yx} binding can be understood in terms of difference in spacing between the origami and gold surface in their respective bound states (bottom left). For \textit{yx} to bind (orange box), the origami must get much closer to the surface than in the case of \textit{xy} (blue box), incurring an increased entropy penalty. Binding of \textit{yx} may also be sterically hindered by the 20~bp long MB reporters that comprise a curtain projecting down from the origami (top left). The \textit{dT28} analyte is a non-complementary sequence and is not expected to trigger any binding event. Experiments were performed on three different chips and are shown as a line for the average and error bars for the standard deviation. (B) The length of the linker tethering the lily pad to the surface has little effect on sensor response: linker lengths of 332, 1037, and 2983~bp all show comparable behavior. Experiments were performed on three different chips and are shown as a line for the average and error bars for the standard deviation. So that overlapping error bars from different experiments can be compared, jitter was added to the \textit{x}-axis.} (C) The functional form of sensor response varies with the concentration of ssDNA analyte. At analyte concentrations up to 500~picomolar (yellow diamonds), sensor response increases monotonically. At nanomolar concentrations (green circles), the system shows saturation effects, and the current  decreases after one hour, potentially due to excess free analyte in solution reopening previously-closed lily pads via DNA strand displacement. ssDNA analyte \textit{xy} added at $t = 0$.}
    \label{fig:2}
\end{figure*}

\subsection*{Chip preparation and electrochemical measurement}
Ultraflat gold chips were fabricated via template-stripping \cite{hegnerUltralargeAtomicallyFlat1993, Weiss2007}.  This approach  had the added benefit of revealing an extremely clean surface upon removal of the template. Two different thiol-modified ssDNA were immobilized on gold surfaces, one that binds the DNA origami linker and one that binds the analyte molecule. The surface was then backfilled with a passivation layer of 6-mercapto-1-hexanol in order to minimize a \added{spurious} current from oxygen reduction\cite{shaverAlkanethiolMonolayerEnd2020}. The chip was incubated with lily pad DNA origami structures and then used as the working electrode in a three electrode cell. 

\subsection*{DNA detection}
\added{In the absence of analyte, SWV of the lily pad- functionalized chip resulted in a baseline voltammogram (an example of which is shown in Fig.~1B, inset; ``Open'' state) with a peak current at $\sim$ -0.27~V, which coincides with the reduction potential of methylene blue\cite{BartonMBpotential1997, xiao_label-free_2005}. We note that the baseline was not observed before incubation of the electrode in the lily pad solution. The stiffness of dsDNA, as measured by its persistence length of $\sim$150~bp\cite{persistence1981,hagerman_flexibility_1988}, is too short for the 1~kbp dsDNA linker to rigidly  hold the lily pads away from the surface. Thus baseline signal, which typically ranges from 0.05 to 0.3~nA,  is likely partially due to thermal fluctuations that bring the MB-modified origami close to the surface. When MB-origami without dsDNA linker are added and rinsed away, we have observed a baseline as high as 0.06~nA. Thus part of the baseline signal may also be attributable to nonspecific binding of the MB-origami part of the lily pad, or free MB-labelled strands. Baselines in the 0.05~nA range are at the limit of our methods to extract meaningful currents, and thus other methods will be necessary to dissect the contributions of tethered yet fluctuating lily pads versus nonspecifically bound MB-origami or MB-labelled strands. }
 
 \added{Similar baselines were} observed for all the SWV presented in this work, and was observed to be stable for over 6~hours and up to four sensor regenerations (Fig.~5). In general, the baseline provides a reference state from which the occurrence of binding events can be inferred.

Upon challenging the sensor with the fully complementary ssDNA analyte \textit{xy}, an increase in SWV signal from baseline was observed (Fig. 1B, inset), resulting in sensorgrams (Fig.~2A, blue curve) similar to those resulting from other real-time surface-bound, DNA hybridization biosensors that use surface plasmon resonance (SPR), biolayer interferometry (BLI) and the E-DNA platform\cite{zhang_electrochemical_2005,Xiao2006,arroyo-curras_electrochemical_2020,xiao_reagentless_2005,arroyo-currasBeakerBodyTranslational2020,lubinFoldingBasedElectrochemicalBiosensors2010}. \added{In addition, this sensorgram demonstrates that, at 500~pM analyte DNA concentration, we observe an increase in current from 0.124~nA to a plateau of 1.58$\pm$0.17~nA ({\em i.e.}, $ \left[ \frac{\mathrm{I_{peak\ for\ saturation}}}{\mathrm{I_{peak\ for\ blank}}} - 1 \right] \times 100 \approx  1,270\%$ gain) after 250~min of target incubation. }For reference, benchmark \textit{optimized} E-DNA ssDNA-detecting sensors displayed gains of 260\% \cite{whiteExploitingBindingInducedChanges2010}; optimized E-AB sensors have reached 430\% \cite{Dauphin-Ducharme2016a}.  Encouraged by these unprecedented gains, we  proceeded to test the effects on sensor response of varying the lily pad and analyte structure.   

For DNA analytes oriented perpendicularly to the surface, like \textit{xy} (Fig.~2A, blue inset), we expected that longer analytes (up to duplex DNA's persistence length) could decrease signal gain by preventing close approach of the 20~bp MB curtain. By switching the positions of the \textit{x} and \textit{y} to yield a sequence (\textit{yx}), the orientation of the analyte was changed to be parallel to the surface (Fig.~2A, orange inset), without significantly affecting predicted thermodynamics. We originally hypothesized that using this orientation could make signal gain independent of length. However, we observed (Fig.~2A, right) slower on-signal kinetics (the rate of signal increase upon addition of analyte) and lower signal at the end of the experiment compared to the \textit{xy} analyte, presumably due to the length of the MB curtain which appears to be a significant steric obstacle for the origami to successfully close. For practical reasons we did not pursue this approach further, but note that shorter MB curtains might recover signal gain and achieve the goal of DNA length-independent signal.

To explore the effect of the lily pad linker length on sensor performance, we synthesized and tested double-stranded DNA linkers of three different lengths: 332, 1037, and 2983~bp (Figure~2B). \added{We challenged these lily pads with 500~pM of \textit{xy} ssDNA analyte; the average behavior of all three sensor types was similar (Fig.~2B; blue, red, and yellow data) in terms of baseline signal (0.17$\pm$0.07~nA), initial on-signal kinetics ($\sim$0.09~nA/min up to 10 min), and endpoint signal (1.49$\pm$0.08~nA).} Given these results, we arbitrarily chose to use the 1037~bp linker for all subsequent experiments.

Figure~2C shows the sensor response to increasing concentrations of DNA analyte. For times less than an hour, both signal and signal kinetics (the rate of signal growth) increase monotonically with concentration. After an hour, two behaviors are observed: (1) signal for analyte concentrations less than 5~nM continues to increase, and (2) signal for an analyte concentration of 5~nM undergoes a surprising decrease. This bifurcation in behavior is reminiscent of the ``high dose hook effect''\cite{HighDoseHookBook2013} which occurs when excess analyte saturates both binders of a sandwich sensor, preventing sandwich formation and decreasing signal. Because the present hook effect emerges over the course of a time-based sensorgram, we term it the ``kinetic hook effect'' (see Supplementary Information Section~6 for discussion).

\begin{figure*}

    \centering
    \includegraphics[width=0.95\textwidth]{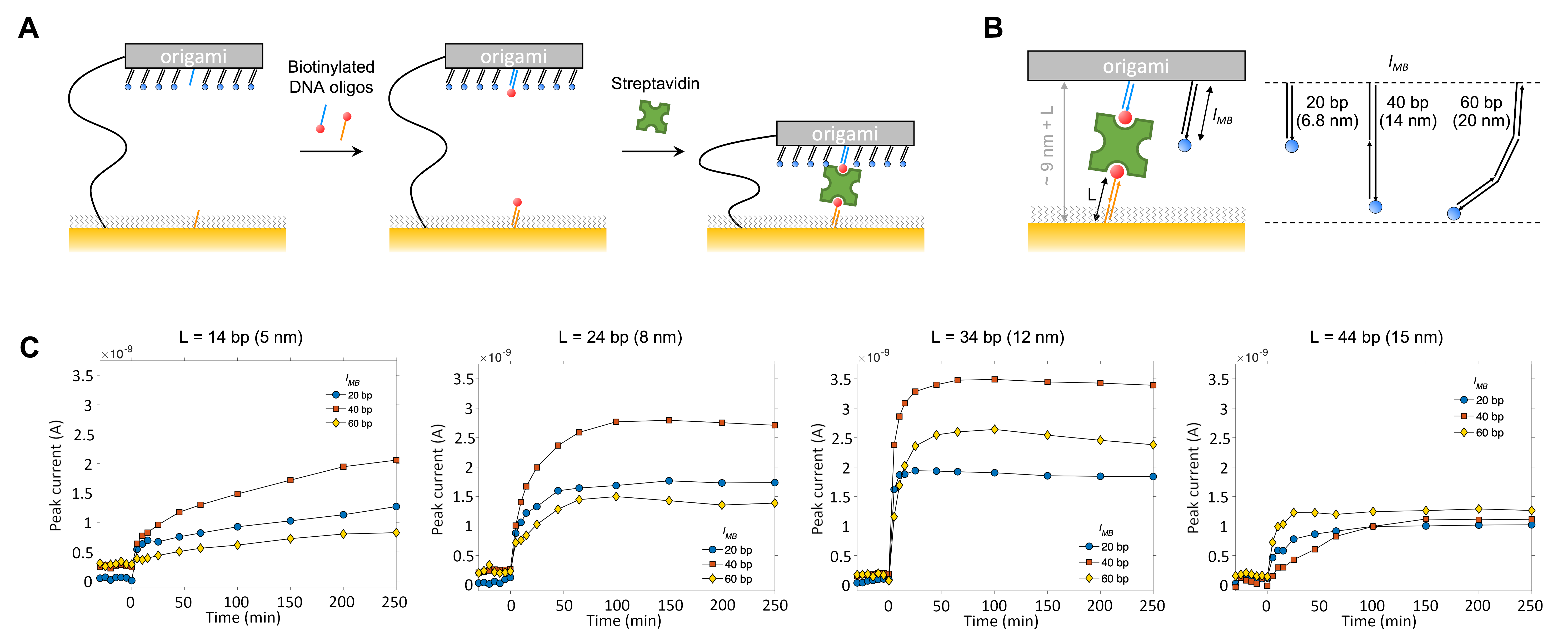}
    \caption{  \textbf{Lily pad sensors for DNA can be readily converted into protein sensors and their response rationally optimized.} (A)  Two biotinylated ssDNA adaptors are hybridized to the sequences \textit{x} and \textit{y} on the origami and on the surface to create a streptavidin sensor. (B) Different lengths for the MB-modified DNA curtain, \textit{$l_{MB}$}, and dsDNA linkers for the surface biotin, \textit{L}, were created to study the effect of molecular design on sensor behavior. (C) SWV sensorgrams for streptavidin (500~pM) detection  with various MB~curtain lengths (20, 40, and 60~bp) on origami and dsDNA linker lengths (14, 24, 34, and 44~bp) on the surface. Different \textit{$l_{MB}$} and \textit{L} give qualitatively different results for both on-signal kinetics and endpoint signal. Longer MB curtains increasingly disturb the closing of the lily pads, but at the same time bring MB closer to the gold surface, increasing peak current; the origami shape itself may have a steric clash with the surface that changes with the length of the analyte-binder complex. The maximum in peak current (40~bp for $L= 14, 24, 34$ and 60~bp for $L=44$) as a function of curtain length is likely set by the tensions between these three effects. Overall, a 40~bp MB curtain and 34~bp surface linker yields the largest signal change.}
    \label{fig:3}
\end{figure*}

\subsection*{Optimizing a Multimeric Protein Detector}
Because of the lily pad's modularity, it is trivial to modify into a detector for multimeric proteins with multiple identical binding sites.  As a proof-of-principle (Fig.~3), we prepared lily pad sensors for detection of a model multimeric protein, streptavidin. We achieved this by simply adding two biotinylated adaptor strands to our basic DNA-sensing chips (Fig.~3A): the first adaptor, sequence  \textit{x}, was complementary to the 14~nt ssDNA tail (\textit{x}$^\prime$) on the origami; the second, sequence \textit{y}, was complementary to the 14~nt oligos (\textit{y}$^\prime$) on the surface. Thus chips with the lily pads used for ssDNA detection were incubated with the biotin-modified adaptors to obtain streptavidin detectors. To optimize sensor design, we measured sensor response as a function of both surface linker length ($L = 14, 24, 34$ or $44$~bp) and MB curtain length ($l_{MB}=$20, 40, or 60~bp). 

Because it is measured before streptavidin addition when the lily pad is nominally open, the baseline was expected to be independent of curtain length. However, the baseline (Fig.~3C) was almost always higher for \mbox{$\textit{l$_{MB}$}=40$}~and 60~bp than for 20~bp (only for $L=44$ was the 40~bp baseline slightly smaller than the 20~bp baseline). Part of the baseline is thus apparently due to transient sticking of the curtain to the surface. To account for longer curtain's enhanced stickiness, we suggest that either (1) longer curtains access MB configurations which are individually stickier, {\em e.g.} with more MB contacting the surface or (2) longer curtains have an increased total number of weakly-sticky MB configurations.

 When streptavidin was added to \mbox{$\textit{l$_{MB}$}=20$}~bp sensors, the SWV signal quickly increased, reaching $\sim$50\% or more of the final signal change within 5~minutes after streptavidin addition. For each $L \leq 34$, comparatively slower on-signal kinetics were observed for \mbox{$\textit{l$_{MB}$}=40$}~bp and 60~bp sensors. The slower kinetics observed for 40~and 60~bp MB curtains (most noticeably for $L=14$) was similar to that observed for detection of the 28~nt DNA reversed sequence, \textit{yx} (Fig.~2A), and we interpret it similarly, {\em i.e.}, we believe that the longer MB curtains sterically hinder sensor closing. However, at $L = 44$, the analyte-binder complex has moved curtains of all lengths sufficiently far from the surface that steric effects are minimal; \mbox{$\textit{l$_{MB}$}=60$}~bp curtains, extending closest to the surface, gave the fastest kinetics.

We expected to see the largest effect of sensor design in peak current endpoints; some experiments showed small decreases in peak current after extended interrogation and so we compared maximum peak currents (MPC) over the 250~min experimental window. Our goal was to optimize signal change and our hypothesis was that trends in sensor performance should be interpretable in terms of the difference $\delta L$ between the total size of the binder-analyte complex and the MB curtains. In an optimal sensor design, the closed lily pad conformation should bring the redox reporters as close as possible to the surface without a steric clash. Absent any tilting of the curtain strands, one might expect to see steric clashes for $\delta L < 0$. In our DNA sensor above, a 28~bp analyte-binder complex (9.5~nm) was used with a 20~bp MB~curtain (6.8~nm) that was slightly shorter (by $\delta L = 2.7$~nm); in this configuration (Fig.~2) the DNA sensor achieved an MPC of 2.75~nA. Table~S6 gives  $\delta L$ and MPCs for all conditions in Fig.~3C. For three of the four conditions (cyan and yellow; Table~S6A) with $\delta L$ most similar to the DNA sensor ($<1.8$~nm different), MPCs were $\geq 75$\% that of the DNA sensor. But a fourth condition was only 45\%. Further, four pairs of conditions having the same $\delta L$ (but different $L$ and $l_{MB}$) had markedly different MPC. This analysis shows that $\delta L$ alone is not a good predictor of MPC.

Instead, observe that for {\em all three curtain lengths} (and thus a wide range of $\delta L$), MPC has a maximum for $L=34$ before steeply dropping off for $L=44$. At this boundary, the ranking of which MB curtain length gives the highest MPC also abruptly changes. Taken together, these observations suggests that there is some other steric effect, which changes sharply between $L=34$ and $L=44$, and which couples to the steric effect of the MB curtain length. Otherwise we would expect a stronger correlation between MPC and $\delta L$ across the boundary. In particular we would expect an increase for the MPC of 60~bp curtains as $L$ changes from 34 to 44, where $\delta L$ goes from 0.92~nm to 4.3~nm. Rather we see a drop in MPC from 2.6~nA to 1.25~nA---for comparison, a large $\delta L$ of 14.5~nm still achieves an MPC of 1.9~nA for 20~bp MB~curtains at $L=34$.

We propose that the origami shapes exhibit either large deformations (either static or dynamic) from a flat disk, on the order of $\sim 21$~nm in height (the size of the analyte-binder complex for $L=34$).
One possible source of deformation is that the MB curtain causes the origami to curl into a U-shaped cross-section that bends up and away from the surface: simulations of origami with MB curtain-like extensions predict such deformations at the 20~nm scale\cite{Sulc2023hairygami}; similar-sized fluctuations of 2D origami have been observed experimentally\cite{floppy2Dorigami2022}. Perhaps for $L=34$ and below, these deformations allow MB curtains of any length to contact the surface, and so increasing $L$ up to~34 decreases steric interference between the MB curtain and the surface. By $L=44$ (corresponding to an analyte-binder complex of 25~nm), all curtain lengths are held too far from the surface for deformations of the origami to bring them into contact. The longest curtains (60~bp, $\sim 20$~nm) give the highest MPC at $L=44$, as they position MB closest to the surface in this ``non-contact'' regime. 

While the dependence of sensor performance on the size of the analyte-binder complex and the MB curtain length is not as simple as we first envisioned, it is nevertheless intelligible. The library of linker and curtain strands we have developed are analyte agnostic. Thus our experiments provide a procedure and map for how sensors for other analytes can be optimized. 
 
Having determined that the optimal (highest MPC) lily pad design for streptavidin detection has a 40~bp MB curtain and \mbox{$L$~$=34$~bp} linker, we sought to assess the linearity of this sensor and measure its sensitivity (LoD). We thus challenged this design with concentrations of streptavidin, ranging from 1~pM to 1~nM (Fig.~4). Signal change was plotted as a function of the streptavidin concentration in Fig.~4 on a semi-log scale. The signal increased monotonically from 2~pM up to 1~nM, with log-linear behavior observed between 5~pM and 1~nM. Using these data, we estimated the limit of detection (LoD) by the conventional \added{method\cite{somasundaram2019,cai2021}}, wherein $\mu_0 + 3\sigma_0$ is the signal at the LoD concentration, $\sigma_0$ is the standard deviation of the blank ([streptavidin] = 0) and $\mu_0$ is the mean of the blank. Here $\mu_0 + 3\sigma_0 = 8.0\%$, where $\mu_0 = -7.0\%$ is sensor baseline in the absence of streptavidin (Fig.~4, red dotted line).  Because [streptavidin] = 1~pM gives a signal change ($17\%$) that is significantly larger than 8.0\%, we infer that LoD~$< 1$~pM. 

As observed with the sensor for the ssDNA analyte \textit{xy}, the gains reported for our optimized streptavidin sensor are unprecedented for a reagentless, unamplified, single-step electrochemical sensor. When challenged with 500~pM streptavidin in bulk solution (Fig.~3C), at steady state measurements past 50~minutes, gains for the 40~bp MB curtain and \mbox{$L$~$=34$~bp} linker system were greater than 1600\%. For experiments where the linearity of the sensor and LoD was determined (Fig.~4), gains at the same streptavidin concentration were calculated to be greater than 500\%. We believe this difference is a function of the experimental conditions. For the steady state measurement, the electrochemical cell was left undisturbed with a large volume of bulk analyte solution (1~mL) and repeatedly interrogated over time. For the linearity/LoD measurement we (1) sought to model a situation with smaller, more practical sample volumes and thus used 10~$\mu$L rather than 1~mL, (2) sought to recycle chips and thus made multiple repeated measurements on the same chip, (3) sought to model an assay in which spuriously-bound sample molecules are washed from the sensor before measurement, and thus the chip was washed with buffer before each baseline and each streptavidin measurement. 

The remarkable size of our sensor's gains becomes clear through comparison with previously-published systems. In comparison to our system, interrogation of other electrochemical streptavidin sensor platforms in buffered solutions gave maximal gains that were {\em at least an order of magnitude smaller} than our maximal lily pad gains. Furthermore, these other platforms' maximal gains were typically reached at significantly higher analyte concentrations. For example, a duplex E-DNA-like sensor displaying biotin gave maximal (in terms of magnitude) gains from -50\% (signal-off) to +50\% (signal-on) at 3~nM streptavidin \cite{StreptavidinSensorForComparison2009}. A modified version of this platform that relies on surface-based steric hindrance, gave -60\% at 100 nM streptavidin \cite{StreptavidinSensorForComparison2015}. A DNA junction-based sensor gave -50\% at 500~nM \cite{StreptavidinSensorForComparison2019}, and a modular, bivalent Y-shaped structure gave -43\% at 100~nM \cite{StreptavidinSensorForComparison2022}. In addition, our sensor's LoD, ~$< 1$~pM, represents an improvement of {\em at least two orders of magnitude} over the LoD of the streptavidin sensors described above.  Given these dramatic gain and LoD improvements over existing sensor designs, we  believe that the lily pad architecture offers a generalizable platform for high-sensitivity, high-gain measurements of analyte concentration. 

\subsection*{Lily pad sensor regeneration}
The biotinylated DNA adapter strands were redesigned to have an additional 5~nt ssDNA toehold (Fig.~5A, Table~S3). This enables their removal via toehold-mediated strand displacement \cite{yurkeDNAfuelledMolecularMachine2000}: a solution containing two 19~nt invading strands (Table~S3) whose sequences are fully complementary to the extended adapter strands is applied to the surface; subsequent strand displacement reexposes the 14~nt DNA-sensing tail (\textit{x}) on the origami and the DNA sensing thiolated ssDNA (\textit{y}) on the surface. Using a 100~nM concentration of applied invading strands, the analyte is displaced in 10~minutes; addition of 500~pM biotinylated adapter strands regenerates the sensor for further interrogation.(Fig.~5B). We conducted four rounds of sensing, displacement and regeneration in this manner, and the sensor responded to streptavidin with similar kinetics after each cycle. 

However, it was observed that this process does not completely restore the sensor to its original state; up to $\sim$5\%~loss in on-signal was seen after each round of regeneration (Fig.~5C). To explain this loss, we suggest that either (1) incomplete strand displacement leaves some streptavidin-biotin complex bound either to the origami or on the surface, blocking analyte binding sites in subsequent sensing rounds or (2) thiolated \textit{y} may desorb from the surface. A third alternative is that (3) MB-strands or even entire lily pads are released with each regeneration; however this is inconsistent with our observation  that the raw baseline sensor signal did not decrease through rounds of regeneration. To achieve greater sensor durability, strategies to increase the efficiency of toehold strand displacement \cite{SimmelYurkeSingh2019StrandDisplacementReview} and improve the robustness \cite{OptimizingElectrochemicalSensorsNetz2023} of gold-bound thiolated sensors to desorption could be applied. 

\begin{figure}
    
    \centering
    \includegraphics[width=3in]{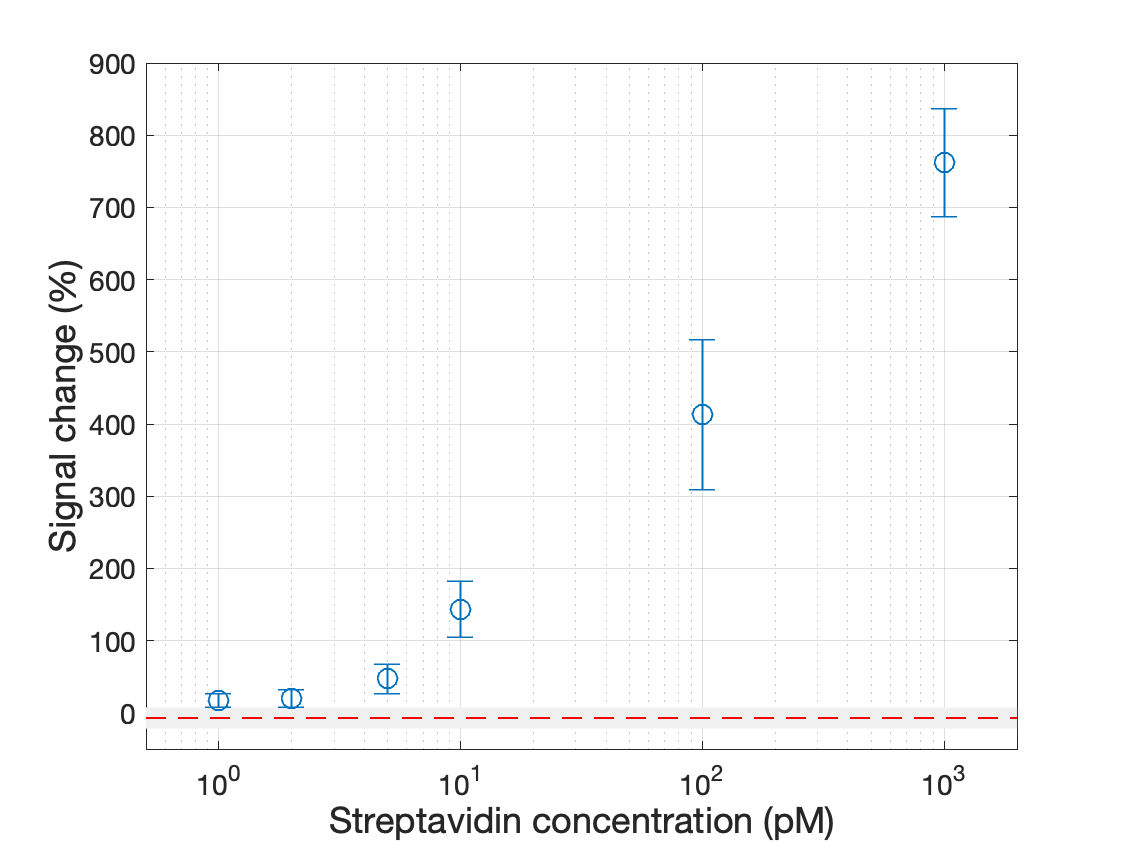}
    \caption{ \textbf{Lily pad sensors can quantify picomolar protein concentrations}. \added{SWV signal changes were measured for lily pads having 40~bp MB curtains and $L=$ 34~bp linkers using six streptavidin concentrations from 1~to 1000~pM and a blank (zero analyte). Mean and standard deviation for each concentration were calculated from replication using five different chips (five biological replicates). On each chip, the off-signal (before sample application) was measured five times and averaged (five technical replicates). After sample application and incubation (one~hour, 34$^\circ$C) the on-signal (endpoint signal) was measured five times and averaged (five technical replicates).} Signal changes in the interval $\mu_0 \pm 3\sigma_0$ are shaded gray, where  $\mu_0$ is the mean of the blank and $\sigma_0$ its standard deviation.  LoD~$< 1$~pM. }
    \label{fig:4}
\end{figure}

\begin{figure*}

    \centering
    \includegraphics[width=0.8\textwidth]{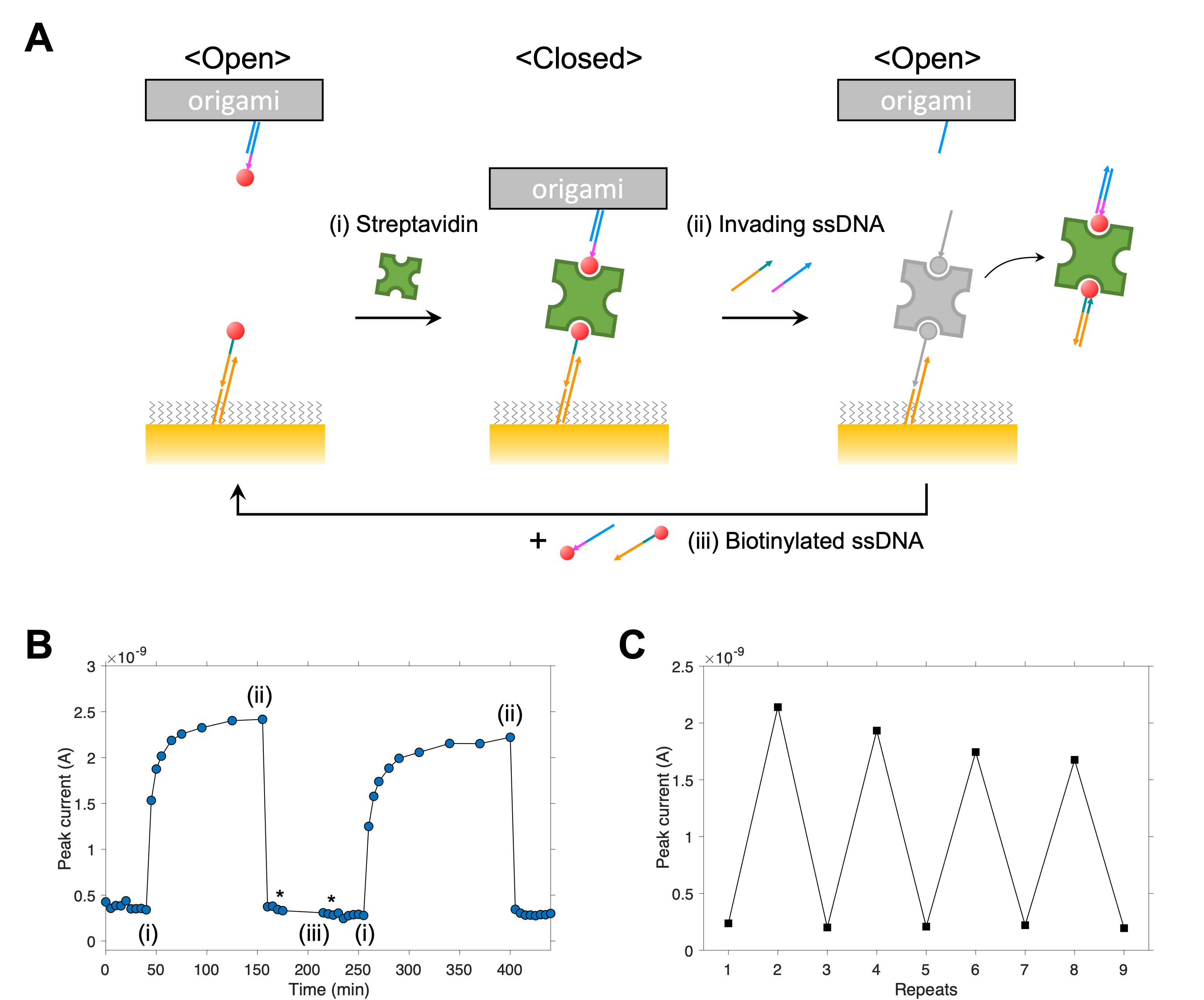}
    \caption{ \textbf{Lily pad sensors with tightly bound analytes can be regenerated multiple times.} (A) Schematic shows a lily pad being closed by streptavidin (i), reopened by ssDNA invaders via strand displacement (ii), and regenerated by the addition of new biotin-modified adaptors (iii). (B) Sensorgram demonstrating two rounds of streptavidin detection where notations are made for buffer washes (*) and the addition of 500~pM streptavidin (i), 100~nM invaders (ii), and 500~pM biotinylated adaptor strands. 
    (C) Changes to SWV signal through four rounds of lily pad regeneration. In each round, streptavidin solution (10~$\mu$L, 500~pM) was reacted for one~hour at 34$^\circ$C  before a measurement was taken (high values). Lily pads with 40~bp MB reporters and $L= 34$~bp were used for (B) and (C).}
    \label{fig:5}
\end{figure*}

\subsection*{PDGF-BB detection}
To further demonstrate the modularity of the lily pad sensor and to test its capability for detecting larger analytes, DNA adapter strands were designed to display an aptamer that binds \added{platelet derived growth factor homodimer of subunit B (PDGF-BB)} (Fig.~6). Since the design of the sensor requires two binding events, one to the origami and one to the surface, PDGF-BB was chosen as the analyte due to its homodimeric nature, which allows for the same aptamer to be used on both the origami and the surface. Surface preparation and origami folding were performed as for previous designs, and adapter strands were added after origami had been tethered to the surface, as with the streptavidin sensor. The two adapters (Fig.~6A) were designed to have sequences \textit{x}$^\prime$ and \textit{y}$^\prime$, followed by a previously reported 36~nt aptamer with an apparent binding affinity to PDGF-BB of 36~fM\cite{vu_oligonucleotide_2017}. The resulting sensor achieved detection of PDGF-BB at concentrations as low as 500~pM (Fig.~6B), with a signal change of $20\%$. Sensor signals were stable after PDGF-BB addition for over 100~mins of serial monitoring every 5~min, highlighting the functional stability of lily pad sensors. The lowest concentration measured is within one order of magnitude of the best sensitivity achieved for previously reported aptamer-based PDGF-BB sensors\cite{laiAptamerBasedElectrochemicalDetection2007}. However, we note that because the goal of these experiments was simply to demonstrate platform modularity, no efforts were made to co-optimize the aptamer and sensor ({\em e.g.} matching aptamer size and MB curtain length); there is therefore, potential to improve the sensitivity of the PDGF-BB version of the lily pad sensor in future work.

\begin{figure}
    
    \centering
    \includegraphics[width=2.6in]{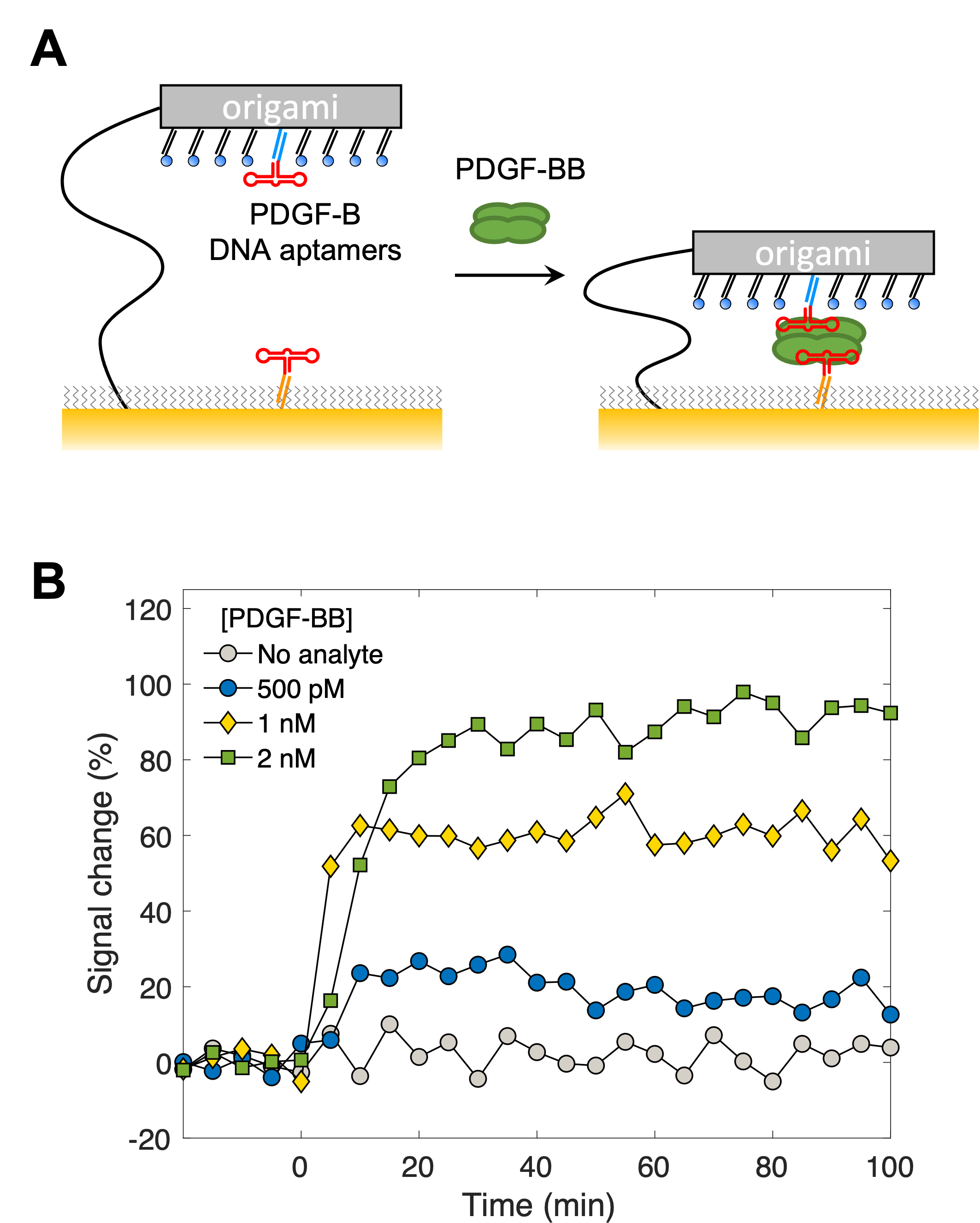}
    \caption{ \textbf{Lily pad sensors can be adapted to detect a clinically-relevant protein biomarker.} (A)  We functionalized lily pad sensors with platelet-derived growth factor (PDGF) binding aptamers. The length and sequence of \textit{x} and \textit{y} adapters were identical to those used in Fig.~3C, as was the sensor preparation and interrogation protocol. (B) Here we show the continuous interrogation of the lily pad PDGF sensor before and after independent addition of three protein concentrations at $t = 0$~min. The root mean square of the baseline ({\em i.e.}, no PDGF) is $\pm 7$\%. Relative to this baseline, the signal to noise ratio (SNR) for each protein challenge is SNR\textsubscript{0.5nM} = 2.8, SNR\textsubscript{1nM} = 8.5 and SNR\textsubscript{2nM} = 13. Lily pad sensors were interrogated every 4~min via SWV, using parameters indicated in the SI.}
    \label{fig:6}
\end{figure}

\section*{Discussion and conclusion}
Using DNA origami, we designed and fabricated a nanodevice---the lily pad---and developed it to create a single-step, reagentless biosensor platform whose modularity enables it to detect arbitrary DNA sequences and proteins through electrochemical measurements. We showed that the modularity of our DNA origami sensor allows the sensing of analytes of varied size and binding properties via simple addition of a few unmodified oligo strands to the base sensor system. The conformational change required for signal is built into the architecture of our sensor, obviating the need to find or engineer binders that undergo a conformational change. Thus the versatility of our sensing platform should only be limited by the ability to functionalize the binding sites of the lily pad. The lily pad can be trivially modified to use the large variety of available aptamers \cite{keefe2010, zhangRecentAdvancesAptamer2019}. Conjugating oligos to other binder classes such as antibodies, antibody fragments\cite{AntibodyFragments2010}, nanobodies\cite{Harmsen2007}, or peptides\cite{wadaDevelopmentNextGenerationPeptide2013} will allow for their facile incorporation into the lily pad. Whenever it is difficult or simply too expensive to get two binders from the same class, hybrid sensors mixing binders from two classes, {\em e.g.} antibodies and aptamers \cite{HybridAntibodyAptamerSandwiches2020}, could be used. Larger binders and/or larger analytes may be accommodated by longer MB curtains or redesign of the lily pad geometry to provide a pocket for the analyte-binder stack. 

In all versions presented here, the lily pad sensors use two binders. In this sensing modality, lily pads can be customized to detect any analytes which are either multimers (for which both the origami and the electrode present the same binder), or have two distinct epitopes (for which the origami and electrode present distinct binders). To access the sensing of analytes with only one available epitope ({\em e.g.} small molecules and some proteins), an appropriate split aptamer\cite{SplitAptamers2020} or aptamer switch\cite{Stojanovic2016CaptureSelex, SohAptamerSwitch2020} could be incorporated into the lily pad. In the case of a split aptamer modality, one half of the aptamer would be attached to the origami, and the other half would be attached to the electrode. Analyte binding to both halves would create a bridge and close the lily pad. In the case of an aptamer switch modality, an aptamer re-engineered to have  an ``antisense'' domain\cite{PlaxcoReingineeringAptamerSensors2010} partially complementary to the analyte-binding region of the aptamer would be attached to the origami.  A sequence complementary to the antisense domain would be attached to the electrode. Upon analyte binding of the aptamer, the antisense domain would be displaced and bind to its surface complement, thus closing the lily pad. Capture SELEX \cite{CaptureSelex2012, Stojanovic2016CaptureSelex} naturally generates switching aptamers with an appropriate antisense domain, and would obviate the need for aptamer re-engineering. Overall, however, split aptamer and aptamer switch modalities would add significant complexity, and can have a poor success rate for some targets \cite{Stojanovic2016CaptureSelex}; thus we expect that aside from the case of small molecules, a sandwich modality will be the preferred format for lily pad sensors.

The modularity of the lily pad provides benefits beyond simply altering the sensor's target specificity: {\em e.g.} we used a library of swappable linkers and curtain strands to optimize sensor performance. Through this approach, we obtained ssDNA (Fig.~2) and streptavidin sensors (Figs.~3 and~4) that can translate picomolar-range changes in analyte concentration to many-fold changes in signal, achieving gains that significantly surpass existing E-DNA/E-AB sensors. The same library could be used to optimize gain and LoD of lily pad sensors that use any of the binders or sensing modalities (sandwich, split aptamer, or aptamer switch) described above. 

Beyond sensor optimization, modularity enables practical chip and sensor reuse. When testing different lily pad designs, we reused chips over ten times by simply treating them with hot water to remove one sensor from the surface linkers, and then adding a new sensor. \added{In separate experiments, our streptavidin sensor regeneration method showed that sensors themselves could be reset to unbound state even with a tightly-bound ligand such as streptavidin, simply by using toehold-mediated strand displacement, and rebinding fresh biotin-displaying adaptor strands.  Furthermore, this approach suggests that sensor specificity could be completely changed ``on the fly'', by exchanging the analyte binding domain using different binder-displaying adaptor strands at every cycle of regeneration, using only these mild, non-denaturing strand displacement reactions.}

 Similar to E-AB sensors, our platform is reagentless and signaling relies on target-binding induced conformational change---it thus has the potential to work in complex biological matrices, {\em in vivo} and ultimately in awake behaving animals.  Achieving this potential will require strategies to increase the lifespan of the electrode and monolayer in serum\cite{DegradationMechanismsSerumAptamerSensors2023}.  Hardening origami to nuclease degradation, low Mg{$^{2+}$} concentrations,  and nonspecific binding {\em in vivo}, while maintaining its structure and functionality, would also be necessary. Such hardening has been achieved by a variety of methods\cite{ChandrasekaranNucleaseResistantDNANanotech2021} including increasing helical packing density, chemical cross-linking, coating with block copolymers and functionalization with unnatural nucleotides and end-groups; these methods should translate readily to the lily pad.
 
Finally, the lily pad design is not limited to electrochemical readout; any readout modality for which signal can be generated by a conformational change should work. With potential modification to the reporter molecules, energy-transfer based fluorescence, surface plasmon resonance, biolayer interferometry, and field effect techniques would all serve as effective readouts for the molecular architecture we describe.

\matmethods{
\subsection* {dsDNA linker preparation}
For synthesis of double-stranded DNA linkers with defined length and single-stranded overhangs with specific sequences, we performed two sequential PCRs, extension and autosticky PCRs\cite{autosticky1999}, using Lambda DNA purchased from Promega (WI) and Taq polymerase from New England Biolabs (MA).
In the extension PCR, each of the DNA primers (sequences in Table~S2), purchased from IDT (CA), consists of a 20~nt long template-binding region that determines the length of the linker and another 20~nt long extension region that determines the terminal sequences of the amplicon added to the sequence from the template DNA (Fig.~S2, top). The amplicons from the first PCR, after purification by agarose gel extraction using Zymoclean Gel DNA Recovery Kits purchased from Zymo Research (CA), were used in the second autosticky PCR as template DNA. Each of the primers used in the autosticky PCR (Fig.~S2, bottom) contains two domains separated by an abasic site, a 20~nt long template-binding region, identical to the extended sequence in the first PCR, and an overhang sequence chosen for binding to either the flat DNA origami or gold surface. The PCR products were purified using the DNA Clean and Concentrator Kit from Zymo Research and stored in 10~mM Tris buffer (pH~7.4) at -20$^\circ$C.

\subsection* {DNA origami design and folding}
The DNA origami we used to prepare the lily pads is derived from a structure used in earlier work \cite{gopinathAbsoluteArbitraryOrientation2021}, which was designed in caDNAno\cite{caDNAno2009} to be a flat, circular shape with a square opening. In that work the square opening gave the design a specific orientation on a microfabricated surface; here the square opening is irrelevant. For the purposes of this work, the design was modified so that extensions of staples on the 5$^\prime$~end would all appear on the same side \added{(Fig.~S8)}. Most staple strands (see Table~S5 for sequences) were ordered from IDT unpurified at 100~$\mu$M in water and stored at -20$^\circ$C. To introduce an analyte binding site near the center for DNA sensing experiments, one staple strand was replaced by a new DNA oligo (IDT, PAGE-purified) that has a 14~nt long 5$^\prime$ extension (shown in cyan in \added{Fig.~S8}). To create 20~bp MB curtains: (1) 
70~out of 234~total staples were extended on their 5$^\prime$ ends with a common 20~nt long single stranded linker and (2) an MB-modified DNA strand (IDT, dual-HPLC purified) with a sequence complementary to the 20~nt linker was hybridized to all 70 extended staples.  

When preparing the lily pads with longer MB curtain lengths (40~or 60~bp) for streptavidin sensing experiments, slightly different schemes were used, as shown at right in Fig.~3B. For 40~bp curtains: (1) a 40~nt MB-modified strand (IDT, Dual-HPLC purified) was hybridized to the standard 20~nt extension, and (2) a single 20~nt ``cap'' ssDNA strand was hybridized to the 3$^\prime$~end of the MB-modified strand to create a fully duplexed curtain. For 60~bp curtains, origami were functionalized with 40~nt extensions that hybridized the 40~nt MB-modified strand via the same 20~nt overlap used for 40~bp curtains. This meant that to create a fully duplexed curtain, both the 20~nt cap described above, a second 20~nt cap were used (where the second cap was hybridized to the section of the staple extension that was proximal to the origami.)

To synthesize origami, we mixed 8094~nt long scaffold strands (p8094 from Tilibit, Germany), with the staple strand mixture, MB-modified DNA oligos (IDT), and the dsDNA linker from PCRs to the final concentrations given in Table~S4 (1:1 scaffold:linker, 5~nM each). One staple on the left side of the origami (yellow orange in \added{Fig.~S8}) was omitted to leave a position on the scaffold at which the dsDNA linker overhang could bind. Throughout this work, we refer to 1$\times$~TAE buffer (Biorad, pH~7.5) with 12.5~mM MgCl$_2$ as ``TAE/Mg''.  100~$\mu$L of scaffold/staple/linker mixture in TAE/Mg was heated to 90$^\circ$C for 5~min and annealed from 90$^\circ$C to 20$^\circ$C at -1$^\circ$C/min. The final concentration of origami, based on the initial scaffold concentration was 5~nM and the solution was then diluted for use to 2~nM in TAE/Mg buffer.

\subsection* {Lily pads assembly on gold surface}
We use the template-stripping (TS) method to prepare ultraflat (see AFM, Fig.~S4) gold surfaces as a substrate for lily pad sensors\cite{hegnerUltralargeAtomicallyFlat1993}. First, a 200~nm gold film was deposited on a 4~inch silicon wafer (University Wafers, MA) using a Labline electron beam evaporator (Kurt J. Lesker Company, PA) at the Kavli Nanoscience Institute Lab at Caltech. The wafer was then cut to $5 \times 8$~mm$^2$ chips by Dynatex GST-150 Scriber/Breaker. A $10\times 10$~mm$^2$ glass coverslip (\#2) was rinsed with acetone, isopropyl alcohol, and water, blown dry with nitrogen, and cleaned by oxygen plasma for 5~min in a PE-50 plasma system (Plasma Etch, NV). About 1~$\mu$L of UV-curable adhesive (Noland, No.61) was applied on a clean glass and a gold chip was placed on top of the adhesive. The adhesive was then cured via a long-wave UV irradiation for 1~hour \cite{Weiss2007}. Right before use, the silicon wafer was pried off the gold/adhesive/glass layers using a razor blade, exposing the ultraflat gold surface on a glass coverslip. To create a well and isolate the reactive area, a silicone gasket (Grace Bio-Labs Press-To-Seal silicone isolator, 2~mm diameter) was glued on top and copper tape was used to form an electrical connection.

Two thiolated DNA oligos were purchased from IDT, one for analyte binding (5$^\prime$-HS-TTTTTAGCTTTGATATCTG-3$^\prime$) and the other for origami-linker tethering on gold (5$^\prime$-CGTAAACCCAGCGTCTTCACCACGATGAATACTCCCACCG- TTTTT-SH-3$^\prime$). In separate tubes, we mixed 1~$\mu$L of 100~$\mu$M thiolated DNA oligos and 1~$\mu$L of 10~mM tris(2-carboxyethyl) phosphine hydrochloride (TCEP, Sigma Aldrich) and incubated for 1~hour at room temperature to reduce the disulfide bonds. Then the solutions were mixed and diluted to 100~nM of each DNA in 1$\times$~PBS buffer (pH~7.4). 20~$\mu$L of the solution was introduced into a silicone gasket well on a freshly prepared TS-gold chip and a Teflon cell (CH instrument, TX) was assembled creating about 1.5~mL of reaction volume that allows three electrode connections. After 1~hour incubation at room temperature, the solution was exchanged with 50~$\mu$L of 10~mM 6-mercapto-1-hexanol (6-MCH, Sigma Aldrich) in 1$\times$~PBS buffer followed by overnight incubation at room temperature for formation of a passivating layer on the gold surface. After this step, the surface presumably had a random distribution of analyte binding and origami linker strands in a 1:1 ratio, with the spaces between these oligos filled by 6-MCH. \added{The density of oligos on these chip surfaces was determined by binding a complementary methylene blue functionalized strand to the chip, and determining total methylene blue occupancy by cyclic voltammetry (Fig.~S5).}

When annealing the lily pad origami structure, we optimized the concentration of the dsDNA linker so that most of the linker was attached to the origami (Fig.~S3). In downstream steps, this allowed us to treat the amount of free linkers as negligible and use the annealed origami mixture without purification. After the passivation step, the 6-MCH solution was removed and 20~$\mu$L of 2~nM MB-modified lily pad solution was added in the reaction well of the silicone gasket and incubated for 30~min at room temperature for origami-linker placement on gold via 40~bp DNA hybridization. Then the gold surface was thoroughly rinsed with TAE/Mg to remove the unbound DNA origami, staple strands, and MB-modified DNA oligos. 

Chips were reused (often more than ten times) by detaching the lily pads and analytes from the surface through rinsing with hot water (65\textdegree~C), by pipetting it directly onto the cell in 100~uL steps for a total of 15--20 times. We confirmed zero MB signal via SWV and cyclic voltammetry measurements after water rinsing.

\subsection* {Electrochemical measurements}
A Metrohm PGSTAT 128N (Netherlands) potentiostat was used for SWV \added{(Fig.~S6)}. After equilibrating the cell at -0.15~V, SWV measurements were performed at a frequency of 10~Hz (see \added{Fig.~S7} for optimization) with an amplitude of 25~mV between -0.15~V and -0.4~V relative to Ag/AgCl reference electrode in 1~mL of TAE/Mg at 34$^\circ$C; temperature was controlled by using a Coy Labs glove box---without strict temperature control sensor response was too variable. SWV voltammograms were recorded every 5~mins for 1~hour before and 5~hours after the addition of analyte DNA, if not otherwise specified. The first hour of measurements before adding analyte is used to set the baseline, off-signal, and measure the fold-increase.
To determine the peak current value a linear baseline is subtracted from the measurement and the MB peak is isolated. This peak is then fit to a Gaussian\cite{gaussianfitCompton2002} and its maximum value is recorded as the peak current \added{(Fig.~S6)}.

\subsection* {Streptavidin detection}
Two biotin-modified strands were ordered from IDT, one to create a surface binding site, complementary to the DNA analyte binding thiolated strand (5'-Biotin-CAGATATCAAAGCT-3'), and the other to create a binding site on origami, complementary to the DNA analyte binding tail (5$^\prime$-CTGAATGGTACGGA-Biotin-3$^\prime$). The DNA sensing chip was incubated for 30~mins with 500~pM of each of these two strands in TAE/Mg at room temperature. This resulted in an $L=14$ sensor. After rinsing with TAE/Mg, SWV measurements were performed for 500~pM streptavidin (Thermo Fischer Scientific) at 34$^\circ$C in the same buffer. 

To create streptavidin sensors with dsDNA surface linkers having a length greater than 14~bp (Fig.~3C), we immobilized one of three different 3$^\prime$-thiolated ssDNAs (10, 20, or 30~nt long) instead of the standard 5$^\prime$-thiolated sequence; we then performed a 6-MCH passivation step. Next, a 30 minute incubation was performed to simultaneously hybridize the lily pads to the surface, and add bridging strands where necessary. In particular, 5~$\mu$M of 24, 34, or 44~nt ssDNA bridging strands were added. The 5$^\prime$ sequence of these strands was complementary to one of the 10, 20, or 30~nt ssDNA already on the surface (as appropriate), so they formed dsDNA complexes proximal to the surface; the 14~nt 3$^\prime$ end of these strands projected ssDNA tails having the sequence \textit{y}$^\prime$ into solution. In the next step where the biotin adapter strands (500~pM) were added, one of the adapters bound to the \textit{y}$^\prime$ tails extending from the surface, forming dsDNA linkers with $L=$ 24, 34, or 44~bp. 

For LoD experiments in Fig.~4, after rinsing off excess biotin-adaptor strands, we measured SWV signal five times before and five times after incubation with the relevant streptavidin solution on the chip; the pre- and post- binding signal for each chip was calculated as the mean of these values. Between the measurements, 10~$\mu$L of varying concentrations of streptavidin samples were introduced into a gasket well on the chip, and sealed with a piece of Parafilm to prevent sample evaporation. After 1~hour incubation at 34$^\circ$C, chips were thoroughly washed with TAE/Mg buffer before the endpoint measurements. This process was repeated for 5 different chips per streptavidin concentration. 

For regeneration experiments in Fig.~5, two 19~nt biotinylated adapters (binding region $+$~5~nt toehold) were used, instead of our standard 14~nt ones. After a streptavidin measurement was finished, the chip was regenerated by adding 100~nM of 19~nt DNA strands that are fully complementary to the adapters, which displace them from the origami and surface sites, returning the chip to the DNA sensing configuration. To finish regeneration of the streptavidin sensor, the chip was incubated with new biotin-modified adapter strands (500~pM, room temperature, 30 min).

\subsection* {PDGF-BB detection}
To create PDGF-BB detecting lily pad sensors, DNA aptamers for PDGF-BB protein (5$^\prime$-CAGGCTACGGCACGTAGAGCATCACCATGATCCTG-3$^\prime$) were ordered from IDT with two 14~nt extensions, one at 3$^\prime$-end for binding to the thiolated DNA strands immobilized on gold surface (5$^\prime$-CAGATATCAAAGCT-3$^\prime$) and the other at 5$^\prime$-end for the single stranded DNA tail on origami (5$^\prime$-CTGAATGGTACGGA-3$^\prime$) that were used for the DNA analyte and streptavidin detection studies (see Table~S3). PDGF-BB protein was ordered from PeproTech, Inc. (NJ). After preparing a standard DNA analyte detecting lily pad sensor, the chip was incubated with 1~nM each of the PDGF-BB aptamers with extensions in TAE/Mg at room temperature. The chip, assembled in a Teflon cell, was rinsed in the same buffer and connected to potentiostat. SWV measurements were performed as described above in 1~mL of TAE/Mg at 30$^\circ$C. After recording SWV voltammograms every 5~mins for 1~hour, 0, 500~pM, 1~nM, or 2~nM of PDGF-BB protein in the same buffer was added.

\subsection* {Data,  Materials,  and  Software  Availability} All  data  needed  to  evaluate  the  conclusions  of  this  study  are  present  in  the  paper  and  its  SI  Appendix.
}

\showmatmethods{} 

\acknow{P.S.L. is grateful for support from Army Research Office (awards W911NF-19-1-0326 and W911NF-231-0283). P.W.K.R. acknowledges support from the Office of Naval Research (awards N00014-18-1-2649 and DURIP N00014-19-1-2341) which provided P.W.K.R, B.J., J.M.S. and M.M.G. with partial support. M.M.G. acknowledge NSF Award No. 2134772 for partial support. J.M.S. acknowledges a fellowship from the Life Sciences Research Foundation supported by Merck Research Laboratories. We thank the Kavli Nanoscience Institute Lab at Caltech for access to fabrication equipment.}

\showacknow{}

\end{document}


\maketitle
\vspace*{-0.3525in}

\vspace*{-0.05in}
\section{Optimizing the number of MB per origami}
\vspace*{-0.05in}
We first tried folding 2D origami with the maximum number of methylene blue (MB) reporters possible, i.e. by extending the 5$^\prime$ end of all 234~staples with linkers complementary to an MB reporter strand. We observed that these heavily-functionalized origami aggregated, presumably due to MB multimerization\cite{braswellEvidenceTrimerizationAqueous1968a,leaistEffectsAggregationCounterion1988} and/or electrostatic interactions between MB and the DNA backbone \cite{vardevanyanMechanismsBindingMethylene2013}. To find the maximum usable number of MB per origami for electrochemical sensors, we prepared 2D origami with different numbers of 20~bp MB reporters (0, 70, 120, and~200) and ran them in an agarose gel. The concentrations of the scaffolds, staple DNA strands, and MB reporter strands are listed in Table~\ref{table:mb_anneal}. 0, 70, 120, and~200 5$^\prime$-extended staple strands were used to create origami with the corresponding number of MB reporters; in each case the rest of the staples did not have single-stranded extensions. Staples were added in a 2$\times$ excess relative to the scaffold and the MB-strands were added at a 1.5$\times$ excess relative to the total concentration of extended staples. Four mixtures were prepared as in Table~\ref{table:mb_anneal}, heated to 90\textdegree C for 5~min and cooled to 20\textdegree~C at -1\textdegree C/min.

\vspace*{-.1in}
\begin{table}[h]
\centering
\begin{tabular}{ll}
\rowcolor[HTML]{A5A5A5} 
Contents           & Concentrations                \\
\rowcolor[HTML]{EDEDED} 
scaffold (p8064)   & 5 				nM                      \\
Staple strands     & 10 				nM (each)              \\
\rowcolor[HTML]{EDEDED} 
MB reporter strand & 0, 				1.05, 1.8, or 3 $\mu$M \\
TAE                & 1$\times$                           \\
\rowcolor[HTML]{EDEDED} 
MgCl$_2$              & 12.5 				mM                  
\end{tabular}
\caption{\vspace*{-0.0in} {\bf Concentrations of components used to fold origami with different numbers of MB reporters.\label{table:mb_anneal}}}
\end{table}

\begin{figure}
    \centering
    \includegraphics[width=0.7\textwidth]{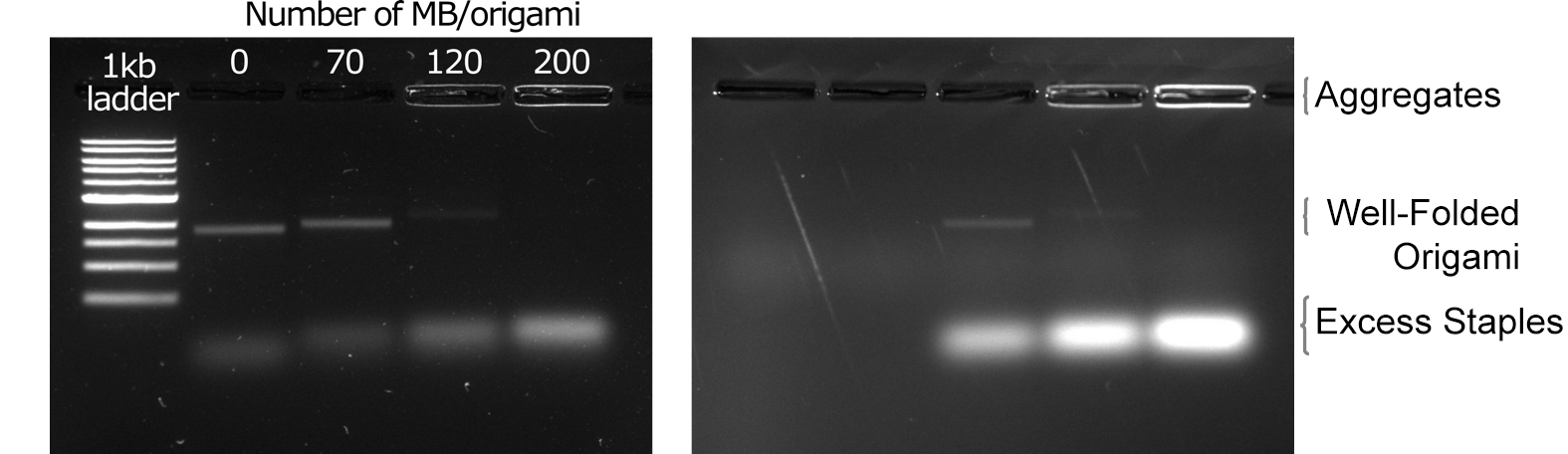}
    \caption{{\bf Gel images for DNA origami with different number of 20~bp MB reporters.} From the left, 1~kb DNA ladder (New England BioLabs), disk origami with zero, 70, 120, and 200 MB reporters, evenly distributed. The gel was imaged with blue (left, for stained DNA) and on red (right, for MB fluorescence) LED illumination. }
    \label{fig:gel_mb}
\end{figure}

\vspace*{-0.1in}
A 1\% agarose gel in TAE buffer with 12.5~mM MgCl\textsubscript{2} was cast with 1$\times$ SYBR-Safe staining dye. 5~$\mu$L of each of 5~nM origami was loaded after being mixed with 1~$\mu$L of 6$\times$ Gel Loading Dye (New England BioLabs). The gel then was run for 1.5~hours at 75~V at 4\textdegree~C in TAE with 12.5~mM MgCl\textsubscript{2} and visualized using a gel imaging system (Syngene G:Box) (Figure~\ref{fig:gel_mb}).

The origami with 0~and 70~MB reporters are seen as single sharp bands under blue LED illumination with 0-MB origami migrating faster than 70-MB origami; this indicates that both the 0-MB and 70-MB origami  are properly folded, monomeric origami structures. On the other hand, both 120-MB and 200-MB functionalized DNA origami are observed as bright bands stuck in the well of the lanes~4 and~5; this suggests that larger numbers of MB reporters (120 and greater) cause origami aggregation. The led us to use 70-MB functionalized 2D origami to construct Lily pads for downstream electrochemical sensing experiments.

\clearpage
\section{Synthesis of dsDNA linkers for Lily pad stalks}
Linkers were created via two rounds of the polymerase chain reaction (PCR). The first (extension) round allowed selection of the length of the final linker (from a 48~kb Lambda DNA template) and added appropriate primers for the next round. 
The second (autosticky \cite{autosticky1999}) round used primers with 40 or 47~nt extensions blocked by abasic sites; polymerase encountering these abasic sites dissociated from the templates without extending to the end of the primers, thus leaving a single-stranded tail at each end of an otherwise dsDNA linker.

\begin{figure}[h]
    \centering
    \includegraphics[width=0.5\textwidth]{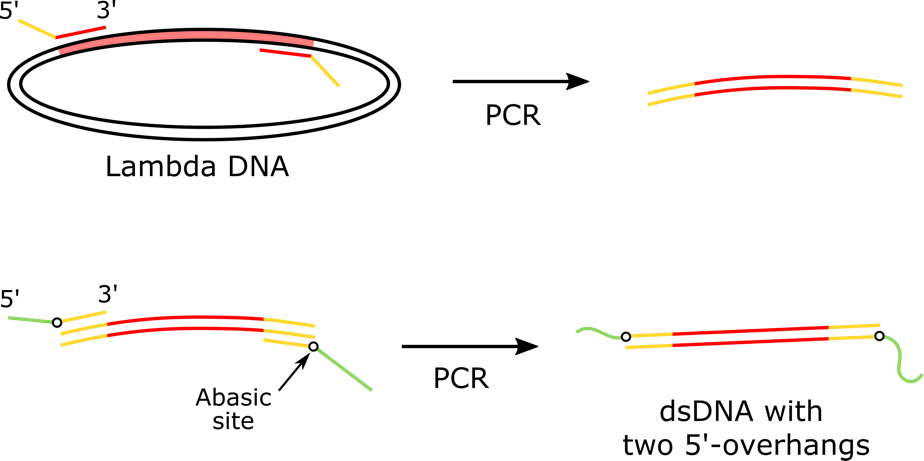}
    \caption{{\bf Synthesis of dsDNA linkers for Lily Pad stalks.} {\bf Top:} a first round of PCR was performed with a pair of primers designed to selectively amplify a subsection of Lambda DNA, having a particular length. {\bf Bottom:} a second round of PCR with primers designed to add single-stranded overhangs was performed. An abasic site between primer sequence and the green single-stranded overhangs caused polymerase to dissociate.}
    \label{fig:pcr_si}
\end{figure}

\begin{table}[h]
\centering
{\scriptsize
\begin{tabular}{ll} 
\rowcolor[HTML]{A5A5A5} 
Name           & Sequences                \\
\rowcolor[HTML]{EDEDED} 
ForPr-extension   & CTACTTAGATTGCCACGCATCTGCCTAGGAATTGGTTAGC \\
RevPr-252bp-extension     & GTAGCATCAGGAATCTGAACGTTTCAGCAGCTACAGTCAG \\
\rowcolor[HTML]{EDEDED} 
RevPr-957bp-extension & GTAGCATCAGGAATCTGAACATGCTCGGAAGGTATGATGC \\
RevPr-2903bp-extension & GTAGCATCAGGAATCTGAACAAGTCCGTGGCTATCTATCG \\
\rowcolor[HTML]{EDEDED} 
ForPr-idSp-DSbind & CACCATCAATATGATATTCAATTTAAATTGTAAACGTTAATATTTTT/idSp/CTACTTAGATTGCCACGCAT \\
RevPr-idSp-40ntOH & CGGTGGGAGTATTCATCGTGGTGAAGACGCTGGGTTTACG/idSp/GTAGCATCAGGAATCTGAAC
\end{tabular}
}
\caption{\vspace*{-0.0in} {{\bf Names and sequences of primers used for synthesis of different lengths of dsDNA linkers for Lily Pad stalks.} \label{table:primers}}}
\end{table}

\begin{table}[h]
\centering
{\scriptsize
\begin{tabular}{ll} 
\rowcolor[HTML]{A5A5A5} 
Name           & Sequences                \\
\rowcolor[HTML]{EDEDED} 
14nt binder 5' extension on origami (top)  & TCCGTACCATTCAG \\
14nt binder 3' anchored on surface (bottom)    & thiol-TTTTT AGCTTTGATATCTG \\
\rowcolor[HTML]{EDEDED} 
28nt DNA analyte \textit{xy} & CTGAATGGTACGGA CAGATATCAAAGCT \\
28nt DNA analyte \textit{yx} & CAGATATCAAAGCT CTGAATGGTACGGA\\
28nt DNA analyte \textit{dT28} & TTTTTTTTTTTTTT TTTTTTTTTTTTTT \\

\rowcolor[HTML]{EDEDED} 
& \\
20~nt 5$^\prime$ staple extension; binds 20~nt MB reporters & GTTGTAGTGGTATGAGGTTG \\
\rowcolor[HTML]{EDEDED} 
20~nt 3$^\prime$ MB-functionalized; for 20~nt MB curtain & CAACCTCATACCACTACAAC-MB \\
40~nt binder for dsDNA linker on surface (bottom) & CGTAAACCCAGCGTCTTCACCACGATGAATACTCCCACCGTTTTT-thiol \\ 
\rowcolor[HTML]{EDEDED} 
Biotin strand with 5nt toehold (top) & CTGAATGGTACGGA-CAACG-biotin \\
Biotin strand with 5nt toehold (bottom)   &  biotin-CCATC-CAGATATCAAAGCT\\
\rowcolor[HTML]{EDEDED} 
Invader strand (top)  & CGTTGTCCGTACCATTCAG \\
Invader strand (bottom)    & AGCTTTGATATCTGGATGG \\
\rowcolor[HTML]{EDEDED} 
PDGF-BB-Aptamer (top)  & CTGAATGGTACGGA CAGGCTACGGCACGTAGAGCATCACCATGATCCTG \\
PDGF-BB-Aptamer (bottom)   & CAGGCTACGGCACGTAGAGCATCACCATGATCCTG CAGATATCAAAGCT \\
\end{tabular}
}
\caption{\vspace*{-0.0in} {{\bf Names and sequences of strands having various functional roles for the Lily Pad sensors.} Strands labeled with ``bottom'' are used on the surface side of the sensor; strands labeled with ``top'' are used on the origami side of the sensor.\label{table:miscellaneous}}}
\end{table}

\clearpage
\section{Lily pad preparation}

Lily pads were prepared by adding dsDNA linkers (``stalks'') with two single-stranded overhangs to the folding reaction of 2D origami with 70 MB reporters. Using the concentration of the linker measured by Nanodrop spectrophotometer and the concentration of scaffold provided by the manufacturer, we performed a titration to find the optimal [linker] to [scaffold] ratio to prepare Lily pad origami with minimal amount of free, unbound dsDNA linkers left so that we can use the annealed origami samples without further purification. Lily pad origami mixtures were prepared as in Table~\ref{table:linker_titration} and annealed from 90\textdegree~C to 20\textdegree~C at -1\textdegree C/min.

\begin{table}[h]
\centering
\begin{tabular}{ll}
\rowcolor[HTML]{A5A5A5} 
Contents                        & Concentrations                                                                 \\
\rowcolor[HTML]{EDEDED} 
scaffold (p8064)                & 5 nM                                                                           \\
Staple strands (70 5’-extended) & 10 nM (each)                                                                   \\
\rowcolor[HTML]{EDEDED} 
MB reporter strand              & 1.05 $\mu$M                                                                    \\
1 kbp linker                    & \begin{tabular}[c]{@{}l@{}}(i) 10 nM\\ (ii) 	5 nM\\ (iii) 	2.5 nM\end{tabular} \\
\rowcolor[HTML]{EDEDED} 
TAE                             & 1$\times$                                                                             \\
MgCl$_2$                           & 12.5 mM                                                                       
\end{tabular}
\caption{{\bf Titration to optimize the amount of 1 kbp dsDNA linker for Lily pad synthesis.\label{table:linker_titration}}}
\end{table}

\begin{figure}[h]
    \centering
    \includegraphics[width=1.0\textwidth]{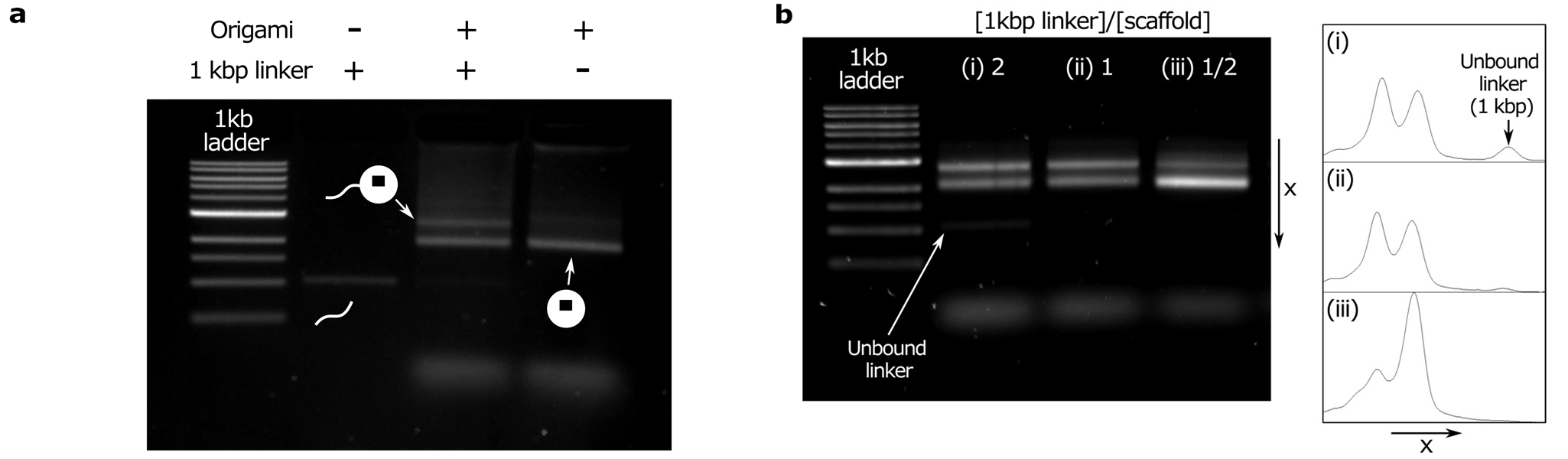}
    \caption{{\bf Gel analyses of Lily pad synthesis.}  ({\bf a}) Gel image for 1~kbp linker, 1~kbp linker + 2D~origami (full Lily pad), and origami only, from left to right. The amount of 1~kbp linker in the lane~2 and lane~3 are same. This gel serves as a reference for the 2D origami and full Lily pad in the gel below. ({\bf b}) Gel image for Lily pad with 70~MB reporters when folded with different amount of 1~kbp linkers. 1\%, SYBR-Safe pre-stained agarose gel was run at 75~V for 90~min at 4\textdegree~C in TAE with 12.5~mM MgCl\textsubscript{2}. [linker] : [scaffold] is reported above each lane.}
    \label{fig:linker_titration}
\end{figure}

Addition of the linker results in appearance of a new, higher molecular weight band in gel as seen in the middle lane in Figure~\ref{fig:linker_titration}a, corresponding to full Lily pads. For comparison, 1~kbp linker and origami only without the linker were run next to the Lily pad lane, the lanes~2 and~4 respectively, where the amounts of the linker in the lanes~2 and~3 are the same, and the amounts of the scaffold (or origami) in the lanes~3 and~4 are matched. 
We found that when [linker] : [scaffold] = 1, a good amount of Lily pad origami structures are formed and the remaining unbound linker is negligible, as seen in the gel image in Figure~\ref{fig:linker_titration}b (left) and in the right panel where the band intensities are plotted for along the DNA migration direction for each sample. For the rest of our Lily pad syntheses we used this 1:1~ratio of linker to scaffold and used the annealed Lily pad origami mixtures without purification, assuming that all the free origami without the linker and the excess staple and MB strands are washed away after lily pads are tethered on gold surface.

\clearpage
\begin{figure}
    \centering
    \includegraphics[width=0.8\textwidth]{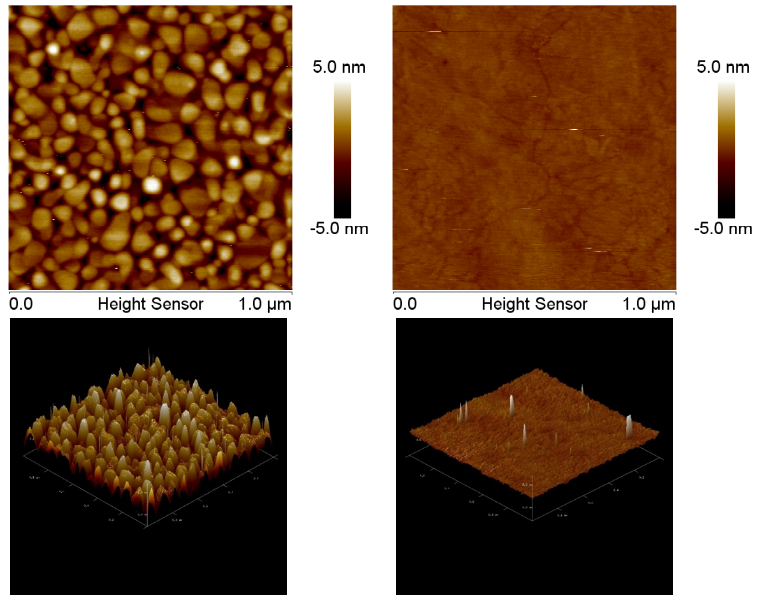}
    \caption{{\bf AFM of gold surfaces.} Electron-beam deposited gold surfaces (AFM at left), commonly used in EDNA~sensors, exhibit 20--100~nm diameter features up to 10~nm in height. EDNA sensors, typically a few nanometers in size, function well on these surfaces as they are much smaller than the features. On the other hand, template-stripped gold surfaces (AFM at right) have a much smoother and flatter profile. Apparent grain boundaries are often hundreds of nanometers apart and roughness is on the order of 1~nm. DNA origami are 100~nm in diameter, and relatively stiff. We reasoned that, to maximize signal in the closed state of the lily pad sensor, we should maximize the number of MB in close proximity to the surface, and we assumed that this would be best accomplished with a template-stripped gold surface; thus these are the surfaces we have used throughout this work. However, we have not compared the performance of the Lily pad on e-beam deposited and template-stripped surfaces.}
    \label{fig:AFM}
\end{figure}

\clearpage

\begin{figure}
    \centering
    \includegraphics[width=0.5\textwidth]{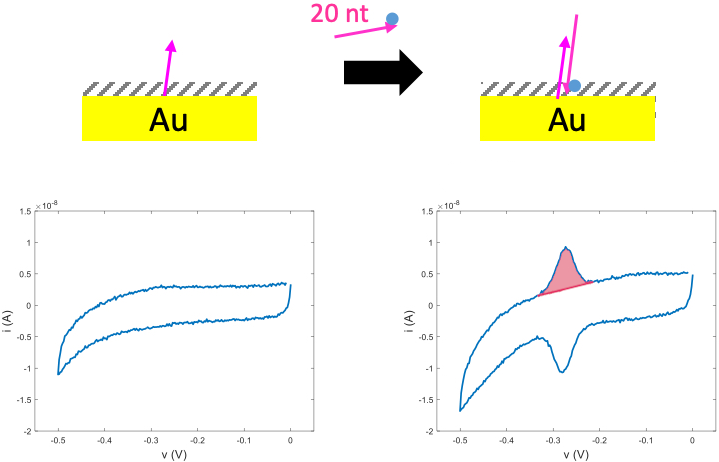}
    \caption{{\bf Measurement of thiolated strand density on gold surfaces.} To estimate the packing density of thiolated strands, 20 ul  a 15 nt thiolated strand at a concentration of 100 nM was deposited on a freshly template-stripped gold surface and incubated for an hour, before washing it off and interrogating the electrode via cyclic voltammetry. The electrode was then incubated with a partially complementary 20 nt strand with a methylene blue modification for 1 h, before rinsing and taking a second cyclic voltammetry measurement. The current peak in this second measurement was then integrated after subtracting the baseline, giving a charge measurement. This arises from the electrons transferred to the electrode from the methylene blue molecules as they are reduced. The charge can then be directly translated into the number of thiolated DNA strands on the surface, assuming full hybridization of the MB-labeled strands. Finally, knowing that template-stripped surfaces are ultra-flat, the surface area is estimated as the area of the silicone gasket. The density was measured on three different gold electrodes using this method and it was estimated to be $3.12 \pm 0.034\ e11\ $strands$/cm^2$.}
    \label{fig:MB_density}
\end{figure}

\begin{figure}
    \centering
    \includegraphics[width=0.8\textwidth]{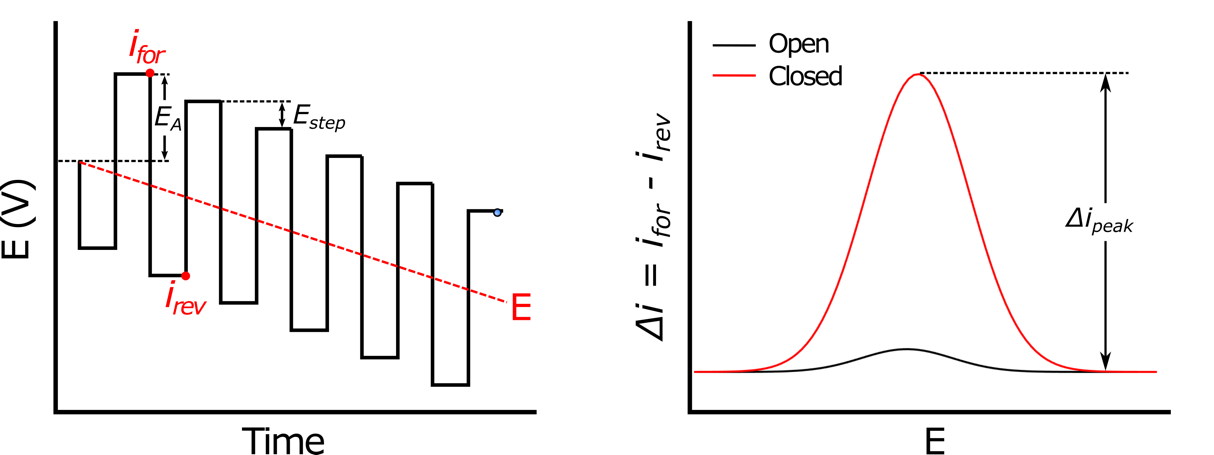}
    \caption{{\bf Square wave voltammetry.} The voltage profile of a single square wave voltammetry measurement is shown at left. A step function is superimposed on a slower voltage ramp and an electric current is measured on both ends ($i_{for}$ and $i_{rev}$) of each step.The difference $\Delta i$ between these two currents, as a function of $E$ is the raw IV Data. To compensate for a tilted, rising baseline at more negative voltage (due to oxygen reduction) the raw IV data is processed as follows. A first approximation to the tilted baseline is made by drawing a line between the current measured at -0.18 V and at -0.40 V of the raw data. This line is then subtracted from the data, to compensate roughly for the tilted baseline. The left and right shoulders of the resulting curve are smoothed with a 7-point moving average, and the maxima of each of these shoulders is found, resulting in a pair of voltages, $V_l$ and $V_r$, which define the voltage range for data processing. The currents in the original IV data at $V_l$ and $V_r$ are used to define a better approximation to the tilted baseline, and this is subtracted from the raw data. This new baseline-subtracted data is fit to a Gaussian curve (between $V_l$ and $V_r$). The peak height is defined as the maximum current of this Gaussian fitted curve.
    The open state of the sensor (black curve) provides a baseline for each experimental condition wherein we expect some fraction of the sensors to be closed (red curve). 
    To determine the sensor signal for a particular experimental condition, the baseline and experimental curves are fit with Gaussians and their peaks are found. 
    The sensor signal or peak current is then defined as~$\Delta i_{peak}$, the difference between the $\Delta i$ for the respective peaks of the two curves.}
    \label{fig:SWV}
\end{figure}

\clearpage
\section{Frequency response of the sensor}
 The frequency response of electrochemical sensors\cite{whiteExploitingBindingInducedChanges2010,Dauphin-Ducharme2017a} is often used to (1) estimate the electron transfer rate of electrochemical sensors and (2) determine the operating frequency of the sensor for which sensor gain and signal-to-noise are optimized. The response of the system is measured by running square wave voltammetry measurements at different frequencies and plotting the peak currents divided by the frequency; this yields response values with units of $A \cdot s$ or charge. Such a plot is known as a Lovri\'{c} plot, or a volcano plot---the name volcano plot derives from the typical shape of these plots, which includes a single, volcano-shaped peak. The frequency at which the peak occurs can be used to derive an electron transfer rate for the sensor\cite{komorsky-lovric_measurements_1995,komorsky-lovric_square-wave_1995}.\\
 
 For typical EDNA systems, in which a short MB-functionalized single-strand is hybridized to its complement on the surface, the peak in the volcano plot (and hence frequency of maximal electron transfer) typically occurs between 10 and 100 Hz. For our Lily pad sensor, the frequency response does not show a well-defined peak over the measured frequency range (1--1000~Hz) for either the open or the closed state. Instead, the response continues to increase as frequency decreases, and any peak likely falls outside of our measurable range. This suggests that, even when Lily pad sensor is closed and the MB are at their smallest distance to the electrode, the MB molecules do not come as close to the surface as they do in other EDNA system. Given the shape of the Lovri\'{c} plot and increased noise at frequencies less than 10 Hz, we chose an operating frequency (10 Hz) for the sensor that gives both good signal gain (between the open and closed states) and relative low noise.

\begin{figure}[h]
    \centering
    \includegraphics[width=0.6\textwidth]{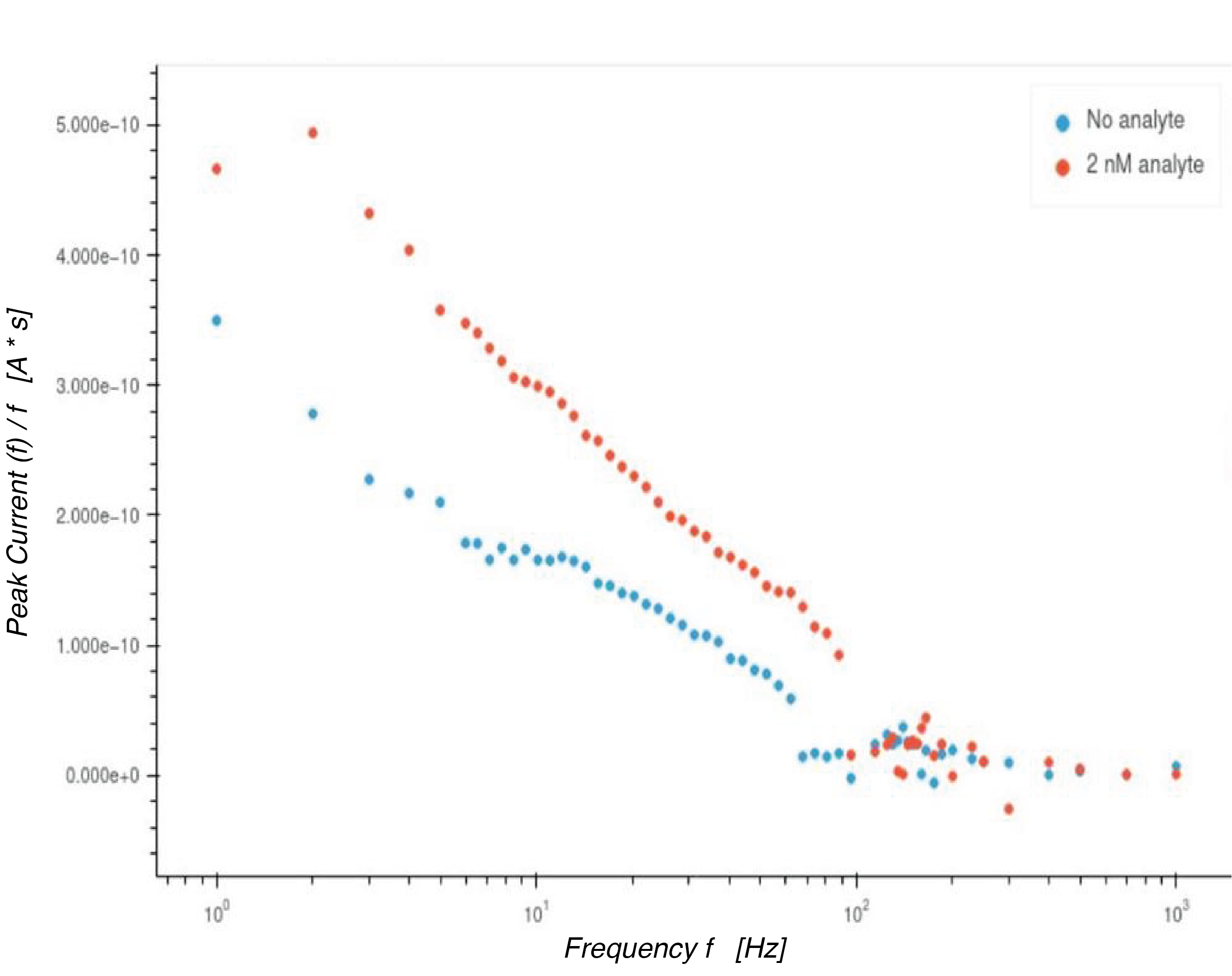}
    \caption{{\bf Frequency response.} The frequency response of the sensor is evaluated using a Lovri\'{c} plot (also known as a volcano plot). The response in $A \cdot s$ calculated as the peak currents at a particular frequency, divided by that frequency. Unlike for usual volcano plots, no clear peak  was observed and we used the plot to select a 10 Hz operating frequency for the sensor, as a compromise between signal gain and signal-to-noise. See Fig.~4 of  Ref.~\cite{komorsky-lovric_square-wave_1995} for an example Lovri\'{c} plot with a clear peak. On the vertical axist that work expresses our  $\mathrm{Peak~Current}(f)/f~[A * s]$ as the dimensionless $(\Delta i_p / f) / \mu C$ where $\Delta i_p$ is our peak current at a particular frequency $f$; note that $A*s$ = $C$. The two plots in Fig.~4 show clear peaks as a function of $f$.}
    \label{fig:lovric}
\end{figure}

\begin{figure}
\section{DNA Origami Design}
    \centering
    \includegraphics[width=1\textwidth]{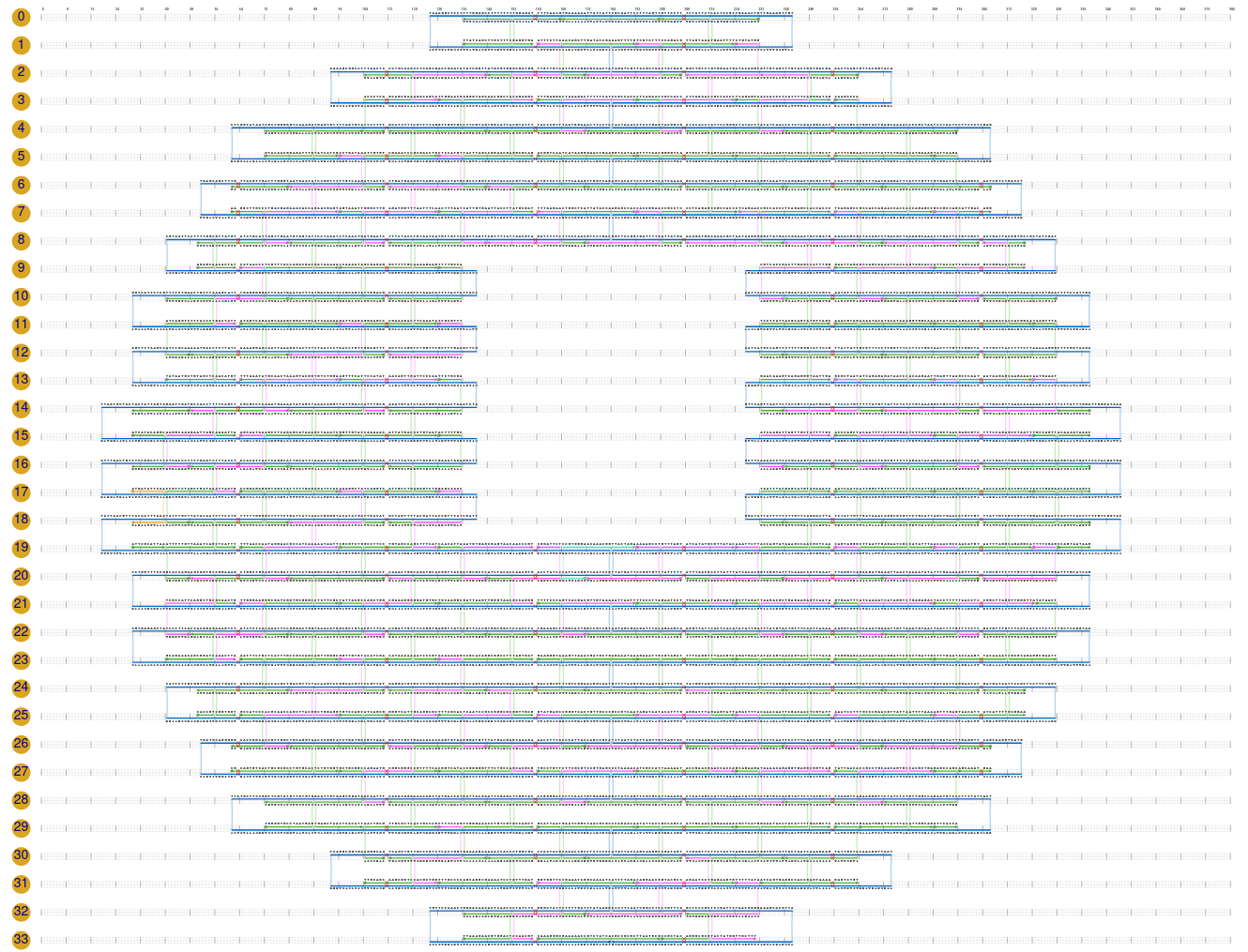}
    \caption{{\bf Design of the DNA origami created using scadnano\cite{doty_et_al:LIPIcs:2020:12962}.} Here the 70-MB variant of the origami used for sensor experiments is depicted.  Strands that were modified with 5$^\prime$ linker extensions, and subsequently bound by MB reporter strands are colored pink. A single staple which was modified with an extension to enable installation of a binder is depicted in cyan (helix 19, at the center). During origami folding, one staple, is omitted from the mixture (the yellow-orange staple in helixes 17 and 18 at center left). This is the position at which one end of the dsDNA linker binds scaffold. The remaining staples are colored green.}
    \label{fig:si_scadnano}
\end{figure}

\newpage

\begin{longtable}{lp{18cm}}
\textbf{Name}                   & \textbf{Sequence}                                                     \\
1{[}136{]}0{[}152{]}   & TTATCAGCTTGCTTTCAAAAAAAGGCTCCAAA                             \\
19{[}64{]}21{[}63{]}   & TTTAACCACCAGCCAGCTTTCCGGCAACTG                               \\
19{[}96{]}21{[}96{]}   & CGCGTCTGGACGACAGTATCGGCCCTCTTCGC                             \\
19{[}128{]}21{[}128{]} & CAACATTACGTAACCGTGCATCTGGGATGTGC                             \\
19{[}328{]}19{[}339{]} & CCAGTAATAAG                                                  \\
22{[}72{]}20{[}88{]}   & GGAATTTGGGCGATCGGTGCGGGCTCAGGAAG                             \\
22{[}136{]}20{[}152{]} & CGGGAACGGATTAAGTTGGGTAACGTGTAGAT                             \\
22{[}200{]}20{[}216{]} & TGCTTCTGAATCCTTGAAAACACGTTAAAT                               \\
21{[}96{]}23{[}96{]}   & TATTACGCTACAGCGCCATGTTTAAGTTGGGC                             \\
21{[}128{]}23{[}128{]} & TGCAAGGCGATAACCTCACCGGAAAATCCCGT                             \\
19{[}29{]}19{[}40{]}   & ATTCGCATTAA                                                  \\
21{[}160{]}23{[}159{]} & TTTTCCCAAGCTTTCAGAGGTGGTAGTGAT                               \\
21{[}224{]}23{[}224{]} & GCTTAGATTCAATATATGTGAGTGTTAACGTC                             \\
21{[}256{]}23{[}255{]} & GTGAATTTATTACCTTTTTTAATTTTTACA                               \\
21{[}288{]}23{[}288{]} & TACCTTTTCAAAATTAATTACATTCCTGATTG                             \\
21{[}321{]}23{[}328{]} & TATAACTAAAGAAGATGATGAAACGCGCAGAGGCGAATT                      \\
23{[}40{]}22{[}56{]}   & GCAGAAACAGCGGATCCTCCGTGG                                     \\
24{[}72{]}22{[}88{]}   & AGCGTGGTCTCATTTGCCGCCAGCCCAGTCCC                             \\
24{[}104{]}22{[}120{]} & ATAACGGACATCGACATAAAAAACAATCGG                               \\
24{[}136{]}22{[}152{]} & TCGCTGGCCCGCACAGGCGGCCTTAGCCGCCA                             \\
24{[}200{]}22{[}216{]} & TCAAAATCGTAGATTTTCAGGTAATAACCT                               \\
21{[}192{]}22{[}200{]} & TTCCCTTAGTAAATCGTCGCTATTACGACGGC                             \\
24{[}232{]}22{[}248{]} & TCTGAATAATACAGTAACAGTACCGGAAACAG                             \\
20{[}296{]}18{[}312{]} & GAAAACTAACATGTAATTTAGGGACAATAA                               \\
20{[}232{]}18{[}248{]} & ATACCGACACCAGTATAAAGCCAATGAACAAG                             \\
15{[}96{]}17{[}96{]}   & GGTTGTACAACCCTCATATATTTTAGATCTAC                             \\
15{[}128{]}17{[}128{]} & ACTTTTGCTTTATTTCAACGCAAGAGAGTCTG                             \\
15{[}256{]}17{[}255{]} & ACCCAGCTTTAGCGAACCTCCCGAGAAACC                               \\
15{[}288{]}17{[}288{]} & ACGCTAACGCTTATCCGGTATTCTTATCATTC                             \\
15{[}320{]}17{[}320{]} & TGCCAGTTCGCCCAATAGCAAGCAACCGCACT                             \\
16{[}328{]}15{[}339{]} & AATCATTACCGACAAAATAAAC                                       \\
18{[}40{]}16{[}56{]}   & AGCAAATACCGTTCTAGCTGATAAGATTCAAA                             \\
18{[}72{]}16{[}88{]}   & AGAAAAGCGGTAGCTATTTTTGAGAAATGCAA                             \\
18{[}104{]}16{[}120{]} & ATGTCAATCAGGTCATTGCCTGGATAAAAA                               \\
20{[}264{]}18{[}280{]} & TCATCTTCCTTAATTGAGAATCGCACGCGCCT                             \\
18{[}232{]}16{[}248{]} & AAAAATAATAATTTACGAGCATGTACTTGCGG                             \\
18{[}296{]}16{[}312{]} & ACAACATGGTATTAAACCAAGTAATCAGAT                               \\
17{[}64{]}19{[}63{]}   & CCGGAGAGCCCAAAAACAGGAAGCTCATTT                               \\
33{[}200{]}32{[}216{]} & CCGCTACAGGGCGCGGCGGGAGC                                      \\
17{[}256{]}19{[}255{]} & AATCAATAACAATAGATAAGTCCCGCTCAA                               \\
17{[}288{]}19{[}288{]} & CAAGAACGGTTCAGCTAATGCAGACATATTTA                             \\
17{[}320{]}19{[}320{]} & CATCGAGAATTCTGTCCAGACGACCAGAGGCA                             \\
18{[}328{]}17{[}339{]} & AAGGTAAAGTAACAAGCAAGCC                                       \\
20{[}40{]}18{[}56{]}   & TCTGGTGCATTTTTGTTAAATCAGATTGTATA                             \\
20{[}104{]}18{[}120{]} & AGGGGACGCCTTCCTGTAGCCAAAACTAGC                               \\
18{[}264{]}16{[}280{]} & GTTTATCAATCGGCTGTCTTTCCTAAGAACGC                             \\
15{[}29{]}16{[}40{]}   & ATCATACAGGCAAAGGCCGGAG                                       \\
24{[}264{]}22{[}280{]} & TCAATATAAACAATAACGGATTCGTAACAATT                             \\
23{[}64{]}25{[}63{]}   & AATTTCTGGCTGGTCTGGTCAGCCGGTGGT                               \\
27{[}256{]}29{[}255{]} & TATTAACATGACCTGAAAGCGTAGACGCTC                               \\
27{[}288{]}29{[}288{]} & GAGCCAGCGGACATTCTGGCCAACATTGGCAG                             \\
27{[}296{]}27{[}307{]} & AGCAAATGAA                                                   \\
29{[}72{]}28{[}88{]}   & TGGGGTGCCTAATGAGCACACAACATACGAGC                             \\
30{[}104{]}28{[}120{]} & TTGCCCTAATTGCGTTGCGCTCCCTGTGTG                               \\
30{[}136{]}28{[}152{]} & TTCTTTTCTCGGGAAACCTGTCGTGCTCGAAT                             \\
30{[}200{]}28{[}216{]} & ACTATCGGCCAGCCATTGCAACGCTATTAG                               \\
30{[}232{]}28{[}248{]} & ATAACATCGAAATACCTACATTTTAGAATACG                             \\
29{[}264{]}28{[}280{]} & GAAATGGATTATTTACAGAGATAG                                     \\
27{[}224{]}29{[}224{]} & GCAGAAGAACAATATTTTTGAATGAGGAAAAA                             \\
29{[}160{]}31{[}159{]} & GCATTAATCGTATTGGGCGCCAGGTTTGAT                               \\
29{[}192{]}30{[}200{]} & ATATTACCGCCTTGCTGGTAATATGGGGAGAG                             \\
29{[}224{]}31{[}224{]} & CGCTCATGACTTGCCTGAGTAGAAGAGAAGTG                             \\
29{[}256{]}31{[}264{]} & AATCGTCTTAGCAATACTTCTTTAGTAAAAGAGTCTGT                       \\
29{[}288{]}29{[}296{]} & ATTCACCA                                                     \\
31{[}104{]}30{[}120{]} & TTGCAGCAAGCGGTCACAGCTGA                                      \\
31{[}128{]}30{[}152{]} & TTTGCCCCAGCAGGCGAAAATCCTGGTGGTTT                             \\
31{[}232{]}30{[}248{]} & ATCAGTGAGGCCACCGGATTAGTA                                     \\
32{[}168{]}31{[}192{]} & ATCAAAAGCGAAATCGGCAAAATCATTTTAGA                             \\
33{[}136{]}32{[}152{]} & TTAAAGAACGTGGACTGGTTGAGTGTTGTTCC                             \\
30{[}168{]}29{[}192{]} & GCGGTTTGGAATCGGCCAACGCGCCCAGAACA                             \\
24{[}296{]}22{[}312{]} & ATATTCCACCAAGTTACAAAATCAAACATC                               \\
27{[}192{]}28{[}200{]} & ATCGCCATTGCGCGAACTGATAGCCTCCTCAC                             \\
27{[}128{]}29{[}128{]} & CGGTCATAATGGTCATAGCTGTTTACTGCCCG                             \\
23{[}160{]}25{[}159{]} & GAAGGGTAACGCGGTCCGTTTTTGGTTGCG                               \\
24{[}168{]}23{[}192{]} & TCCGGCAAAAGTTAAACGATGCTGAGAAATAA                             \\
23{[}192{]}24{[}200{]} & AGAAATTGTATTTGCACGTAAAACATTGCCGT                             \\
23{[}224{]}25{[}224{]} & AGATGAATATGGAAGGGTTAGAACGAAGTATT                             \\
23{[}256{]}25{[}255{]} & TCGGGAGAATCCTGATTGTTTGGTCGTATT                               \\
23{[}288{]}25{[}288{]} & CTTTGAATTGATTATCAGATGATGATTTTAAA                             \\
25{[}50{]}24{[}56{]}   & CACCGTAGCAAC                                                 \\
26{[}104{]}24{[}120{]} & GTTAACGATCAGCGGGGTCATTACATCCTC                               \\
26{[}232{]}24{[}248{]} & ATTGAGGACAAACAATTCGACAACATTATACT                             \\
27{[}160{]}29{[}159{]} & TGCCTGTTATCCCCGGGTACCGAGCCAGCT                               \\
26{[}264{]}24{[}280{]} & TGGTCAGTTGCCCGAACGTTATTAGCAATTCA                             \\
24{[}312{]}25{[}318{]} & GCGGAAGCGGAA                                                 \\
25{[}64{]}26{[}72{]}   & GCCATCCCTGCCGGTGCC                                           \\
25{[}96{]}27{[}96{]}   & TGTCCAGCGCATCAGATGCCGGGTTCTGTGGT                             \\
25{[}128{]}27{[}128{]} & TTTCGCACTTACACTGGTGTGTTCCGTTTTCA                             \\
26{[}168{]}25{[}192{]} & GTAATGGGCGGGTCACTGTTGCCCGCCGTCAA                             \\
27{[}61{]}26{[}88{]}   & CGCAGTGTCACTGCGCGCCTGTGCACTACCTGCA                           \\
28{[}136{]}26{[}152{]} & TCGTAATCCCGGGGGTTTCTGCCATCGTCATA                             \\
28{[}200{]}26{[}216{]} & TCTTTAATAAAAATACCGAACGCTTTAGGA                               \\
27{[}96{]}29{[}104{]}  & GCTGCGGCTATCCGCTCACAATTCTGAGCTAACTCACATT                     \\
26{[}296{]}24{[}312{]} & CTGAACCTCATAACATTATCATTTTTTATCATC                            \\
14{[}328{]}14{[}339{]} & TCCAAATAAGA                                                  \\
33{[}160{]}33{[}200{]} & CAAAGGGCGAAAAACCGTCTATCACGCCGCGCTTAATGCG                     \\
8{[}168{]}7{[}192{]}   & GTTGGGAACTGGCTCATTATACCAAGAGCCAC                             \\
10{[}104{]}8{[}120{]}  & CGTCCAAGCGAGAGGCTTTTGCGGTAGAAA                               \\
6{[}200{]}4{[}216{]}   & TTTTGATGGGGTCAGTGCCTTGAGGTGTAT                               \\
6{[}136{]}4{[}152{]}   & ACCAGGCGTACTTAGCCGGAACGATAATGCCA                             \\
6{[}104{]}4{[}120{]}   & GAGTAATATTGTGTCGAAATCCAAAGAGGC                               \\
8{[}312{]}9{[}318{]}   & GCACCGCGTCAC                                                 \\
4{[}88{]}5{[}96{]}     & TACACTAAAACACTCATATCATCG                                     \\
5{[}72{]}4{[}88{]}     & GTACAACGGAGATTTGTCTTTGACCCCCAGCG                             \\
3{[}256{]}5{[}255{]}   & CAAGCCCAGGCGGATAAGTGCCGCCTGCCT                               \\
3{[}224{]}5{[}224{]}   & TCAGAGCCAAGTATAGCCCGGAATAGTAACAG                             \\
9{[}96{]}11{[}96{]}    & CCAAAATATACTGCGGAATCGTCAATCAAAAA                             \\
3{[}192{]}4{[}200{]}   & CTCAGAACCTCAGGAGGTTTAGTAGAGGAAGT                             \\
9{[}128{]}11{[}128{]}  & TTTTGCCATAGTAAAATGTTTAGATTCAAATA                             \\
5{[}192{]}6{[}200{]}   & GTTTTAACGATACAGGAGTGTACTCTGACCAA                             \\
3{[}160{]}5{[}159{]}   & TTTGAGGAAACGGGTAAAATACGGGCGCAG                               \\
3{[}128{]}5{[}128{]}   & TCGGAACGGCACCAACCTAAAACGGCGACCTG                             \\
10{[}72{]}8{[}88{]}    & CATTGAATTCGTTTACCAGACGACATACCACA                             \\
2{[}248{]}3{[}255{]}   & ACTGAGTTTCGTCACGGGATAG                                       \\
9{[}50{]}8{[}56{]}     & AGAGCAGCCAAA                                                 \\
6{[}232{]}4{[}248{]}   & AATTTACCTAAACAGTTAATGCCCTCGAGAGG                             \\
5{[}224{]}7{[}225{]}   & TGCCCGTAGTTCCAGTAAGCGTCACCGCCACCC                            \\
5{[}256{]}7{[}255{]}   & ATTTCGGACAGAATGGAAAGCGCACCACCC                               \\
5{[}288{]}7{[}288{]}   & TATTAAGAATATTCACAAACAAATCCAGAGCC                             \\
5{[}160{]}7{[}159{]}   & ACGGTCAAGAGGACAGATGAACGATGCGAT                               \\
6{[}61{]}6{[}80{]}     & ACCCAAATCAACGTAACA                                           \\
8{[}200{]}6{[}216{]}   & TCATAATCCGCCTCCCTCAGAGTACATGGC                               \\
8{[}232{]}6{[}248{]}   & TCATAGCCCGCCACCCTCAGAGCCAGTCTCTG                             \\
8{[}264{]}6{[}280{]}   & GACTGTAGGCCACCAGAACCACCAAAATCCTC                             \\
6{[}296{]}6{[}307{]}   & ATTGGCCTTG                                                   \\
7{[}61{]}9{[}63{]}     & AAGGCTTGCCATGCAGATACATAACACACTAT                             \\
7{[}96{]}9{[}96{]}     & AGTAAATTCAGTTGAGATTTAGGAGATAAAAA                             \\
7{[}128{]}9{[}128{]}   & ACTTTAATGAACAACATTATTACAAAAAGAAG                             \\
4{[}280{]}5{[}288{]}   & GATTAGGATTAGCGGGCATGAAAG                                     \\
16{[}104{]}14{[}120{]} & TTTTTAGCAAAAACATTATGACATACATTT                               \\
6{[}264{]}4{[}280{]}   & ATTAAAGCACCTATTATTCTGAAAGTTTTGCTCAGTACCA                     \\
8{[}296{]}7{[}307{]}   & AGCGACACATTGACAGG                                            \\
4{[}136{]}2{[}152{]}   & CTACGAAGAGGGTAGCAACGGCTACCACGCAT                             \\
9{[}256{]}11{[}255{]}  & GAATTAGATAAATATTGACGGAATGTTAGC                               \\
6{[}168{]}5{[}192{]}   & CTTTGAAATCATAAGGGAACCGAAGGTAATAA                             \\
14{[}264{]}12{[}280{]} & ACATAAAAATCAGAGAGATAACCCGAAACCGA                             \\
14{[}296{]}12{[}312{]} & AATGAAAGCCCAATAATAAGAGTAAGCAGA                               \\
13{[}40{]}12{[}56{]}   & TATAATGCTGTAGCTCGGTCATTT                                     \\
11{[}320{]}13{[}320{]} & TAAGTTTACCCTTTTTAAGAAAAGCAAGAAAC                             \\
13{[}64{]}15{[}63{]}   & TTTAAATAGGGGCGCGAGCTGAAAAAATTA                               \\
2{[}168{]}1{[}192{]}   & CCGACAATAACAGCTTGATACCGAAAGTTTTG                             \\
11{[}288{]}13{[}288{]} & CATATAAAACAAAGTTACCAGAAGACAAGAAT                             \\
11{[}256{]}13{[}255{]} & AAACGTAGAATAATAACGGAATATAATTGA                               \\
13{[}96{]}15{[}96{]}   & AGTTTCATGGTCAATAACCTGTTTGCTAAATC                             \\
13{[}128{]}15{[}128{]} & TTCTGCGATTAGTTTGACCATTAGCCTGTAAT                             \\
2{[}200{]}0{[}216{]}   & CTCATAGCCAGACGTTAGTAAAAGAATAGAAAGGAAC                        \\
11{[}64{]}13{[}63{]}   & ATAGTCAGCTCCTTTTGATAAGAAACATGT                               \\
4{[}104{]}2{[}120{]}   & AAAAGAAACCCTCAGCAGCGAACTTGCAGGGAGTTAA                        \\
10{[}312{]}11{[}320{]} & TTCATATGGTTTACCACCACGGAA                                     \\
12{[}296{]}10{[}312{]} & TAGCCGAAGAAACGCAAAGACAGCGCCAAA                               \\
12{[}264{]}10{[}280{]} & GGAAACGCAAAATACATACATAAAATTGAGGG                             \\
12{[}232{]}10{[}248{]} & AACTGGCATCCTTATTACGCAGTAATTATTCA                             \\
14{[}29{]}14{[}48{]}   & ATCAATTCTACTAATAGTA                                          \\
12{[}104{]}10{[}120{]} & AAACTCCGGAAGCCCGAAAGACCTGGATAG                               \\
12{[}72{]}10{[}88{]}   & TTTAATTGAAGCAAAGCGGATTGCTAAATATT                             \\
12{[}40{]}10{[}56{]}   & TTGCGGATATCAGGTCTTTACCCTTAAACAGTTCAGAAAA                     \\
9{[}288{]}11{[}288{]}  & ATTACCATGGGCGACATTCAACCGGGTGGCAA                             \\
0{[}168{]}0{[}200{]}   & AACTAAAGGAATTGCGAATAATAATTTTTTCA                             \\
16{[}72{]}14{[}88{]}   & TGCCTGAGAGCCTCAGAGCATAAAAGCTATAT                             \\
0{[}152{]}1{[}159{]}   & CGTTGAAAATCTCCAGAGGTGA                                       \\
14{[}232{]}12{[}248{]} & GAATTAACGAACAAAGTCAGAGGGCCCAAAAG                             \\
0{[}216{]}1{[}224{]}   & CAGTTTCAGCGGAGTGTGAATTTT                                     \\
20{[}168{]}19{[}192{]} & CCGTAATGCCGTGGGAACAAACGGTTACTAGA                             \\
17{[}29{]}18{[}40{]}   & ATGATATTCAATTTAAATTGTA                                       \\
28{[}104{]}26{[}120{]}\_MB & GTTGTAGTGGTATGAGGTTGAAATTGTCAGAATGCGGCGGGCAGCAAATC           \\
29{[}128{]}31{[}128{]}\_MB & GTTGTAGTGGTATGAGGTTGCTTTCCAGACCAGTGAGACGGGCACACGCTGG         \\
31{[}224{]}33{[}231{]}\_MB & GTTGTAGTGGTATGAGGTTGTTTTTATACCTCGTTAGAATCAGATACTATGGTTGCTTT  \\
4{[}200{]}2{[}216{]}\_MB   & GTTGTAGTGGTATGAGGTTGCACCGTACGCCACCCTCAGAACAGACAGCC           \\
31{[}192{]}32{[}200{]}\_MB & GTTGTAGTGGTATGAGGTTGCAGGAACGGAGGCCGATTAAAGGGCCTTATAA         \\
5{[}128{]}7{[}128{]}\_MB   & GTTGTAGTGGTATGAGGTTGCTCCATGTCATAGGCTGGCTGACCTAATTTCA         \\
4{[}232{]}2{[}248{]}\_MB   & GTTGTAGTGGTATGAGGTTGGTTGATATACCACCCTCATTTTCACAGTACAAACTACAAC \\
25{[}160{]}27{[}159{]}\_MB & GTTGTAGTGGTATGAGGTTGGTATGAGCTAAAGGTTTCTTTGCGCACGCG           \\
5{[}96{]}7{[}96{]}\_MB     & GTTGTAGTGGTATGAGGTTGCCTGATAACTTGACAAGAACCGGAGAACGAGT         \\
25{[}192{]}26{[}200{]}\_MB & GTTGTAGTGGTATGAGGTTGTAGATAATCAACTAATAGATTAGATGCGGCTG         \\
25{[}224{]}27{[}224{]}\_MB & GTTGTAGTGGTATGAGGTTGAGACTTTAAGGTTATCTAAAATATAACCACCA         \\
4{[}168{]}3{[}192{]}\_MB   & GTTGTAGTGGTATGAGGTTGTTCCATTACTAAAGACTTTTTCATCCGCCACC         \\
31{[}160{]}33{[}159{]}\_MB & GTTGTAGTGGTATGAGGTTGGGTGGTTCAATAGCCCGAGATAGCCAACGT           \\
1{[}224{]}3{[}224{]}\_MB   & GTTGTAGTGGTATGAGGTTGCTGTATGGGCCTGTAGCATTCCACCGCCACCC         \\
32{[}200{]}30{[}216{]}\_MB & GTTGTAGTGGTATGAGGTTGTAAACAGGTACGCCAGAATCCTGAACTCAA           \\
25{[}256{]}27{[}255{]}\_MB & GTTGTAGTGGTATGAGGTTGAAATCCTTTGGCAAATCAACAGTCGGTCAG           \\
25{[}288{]}27{[}288{]}\_MB & GTTGTAGTGGTATGAGGTTGAGTTTGAGAATATCAAACCCTCAACACGCTGA         \\
28{[}168{]}27{[}192{]}\_MB & GTTGTAGTGGTATGAGGTTGAGTTGAGGCTTCGCGTCCGTGAGCCCTAAAAC         \\
1{[}192{]}2{[}200{]}\_MB   & GTTGTAGTGGTATGAGGTTGTCGTCTTTTTAGCGTAACGATCTATAGTTGCG         \\
28{[}264{]}26{[}280{]}\_MB & GTTGTAGTGGTATGAGGTTGAACCCTTCCCGCCTGCAACAGTGCTCAATATC         \\
1{[}160{]}3{[}159{]}\_MB   & GTTGTAGTGGTATGAGGTTGATTTCTTAGACAACAACCATCGCCAGAGGC           \\
2{[}120{]}3{[}128{]}\_MB   & GTTGTAGTGGTATGAGGTTGAACCGATATATTCGGTCGCTGAGGAGACAGCA         \\
28{[}232{]}26{[}248{]}\_MB & GTTGTAGTGGTATGAGGTTGTGGCACAGTAAAACAGAGGTGAGGTGAAAGGA         \\
16{[}232{]}14{[}248{]}\_MB & GTTGTAGTGGTATGAGGTTGGAGGTTTTATCAAGATTAGTTGCTAGACGGGA         \\
26{[}136{]}24{[}152{]}\_MB & GTTGTAGTGGTATGAGGTTGAACATCCCTCAATCCGCCGGGCGCTCGTCTCG         \\
19{[}256{]}21{[}255{]}\_MB & GTTGTAGTGGTATGAGGTTGCAGTAGGGTGACCTAAATTTAATGTCAATA           \\
19{[}224{]}21{[}224{]}\_MB & GTTGTAGTGGTATGAGGTTGAAATTCTTCGTGTGATAAATAAGGTAGCGATA         \\
20{[}200{]}19{[}224{]}\_MB & GTTGTAGTGGTATGAGGTTGAAGAATAGTTTAGTATCATATGCGTTATAC           \\
19{[}192{]}20{[}200{]}\_MB & GTTGTAGTGGTATGAGGTTGAAAAGCCTAACACCGGAATCATAACGGATTGA         \\
19{[}160{]}21{[}159{]}\_MB & GTTGTAGTGGTATGAGGTTGCGGATTCTGGATAGGTCACGTTGGCCAGGG           \\
20{[}136{]}19{[}159{]}\_MB & GTTGTAGTGGTATGAGGTTGGGGCGCATAATGTGAGCGAGTAACAACCCGT          \\
20{[}72{]}18{[}88{]}\_MB   & GTTGTAGTGGTATGAGGTTGATCGCACTATAGGAACGCCATCAATGATAATC         \\
11{[}96{]}13{[}96{]}\_MB   & GTTGTAGTGGTATGAGGTTGGATTAAGAAACAGGTCAGGATTAGTGTCTGGA         \\
11{[}128{]}13{[}128{]}\_MB & GTTGTAGTGGTATGAGGTTGTCGCGTTTTTCAAAGCGAACCAGAATTCCCAA         \\
17{[}128{]}19{[}128{]}\_MB & GTTGTAGTGGTATGAGGTTGGAGCAAACATGAACGGTAATCGTAGCTTTCAT         \\
14{[}72{]}12{[}88{]}\_MB   & GTTGTAGTGGTATGAGGTTGTTTCATTTTGCAACTAAAGTACGGAGAGTACC         \\
14{[}104{]}12{[}120{]}\_MB & GTTGTAGTGGTATGAGGTTGCGCAAATTCCATATAACAGTTGCCGGAAGC           \\
13{[}256{]}15{[}255{]}\_MB & GTTGTAGTGGTATGAGGTTGGCGCTAATACAGGGAAGCGCATTATTTTGC           \\
13{[}288{]}15{[}288{]}\_MB & GTTGTAGTGGTATGAGGTTGTGAGTTAAATAGCAGCCTTTACAGTCTTACCA         \\
13{[}320{]}15{[}320{]}\_MB & GTTGTAGTGGTATGAGGTTGAATGAAATAACGATTTTTTGTTTAGCCTAATT         \\
15{[}64{]}17{[}63{]}\_MB   & GTTGTAGTGGTATGAGGTTGAGCAATAATAATGTGTAGGTAAAATTAATG           \\
16{[}40{]}14{[}56{]}\_MB   & GTTGTAGTGGTATGAGGTTGAGGGTGAGAAGGCAAAGAATTAGCAAGGTGGC         \\
16{[}296{]}14{[}312{]}\_MB & GTTGTAGTGGTATGAGGTTGATAGAAGGAGCGTCTTTCCAGAACGTCAAA           \\
16{[}264{]}14{[}280{]}\_MB & GTTGTAGTGGTATGAGGTTGGAGGCGTTACAATTTTATCCTGAAAGAGAATA         \\
19{[}288{]}21{[}288{]}\_MB & GTTGTAGTGGTATGAGGTTGACAACGCCTTTTCAAATATATTTTTGAGAGAC         \\
19{[}320{]}21{[}321{]}\_MB & GTTGTAGTGGTATGAGGTTGTTTTCGAGCCAATCGCAAGACAAAGTTGGGTTA        \\
22{[}40{]}20{[}56{]}\_MB   & GTTGTAGTGGTATGAGGTTGTGAAGGGATCGCCATTCAGGCTGCGCACCGCT         \\
22{[}104{]}20{[}120{]}\_MB & GTTGTAGTGGTATGAGGTTGCGAAACGCAGCTGGCGAAAGGGCCAGTTTG           \\
26{[}72{]}24{[}88{]}\_MB   & GTTGTAGTGGTATGAGGTTGGCCAGCGGACGCAACCAGCTTACGAGAACGTC         \\
8{[}72{]}6{[}88{]}\_MB     & GTTGTAGTGGTATGAGGTTGTTCAACTACTGACGAGAAACACCATATTCATT         \\
8{[}104{]}6{[}120{]}\_MB   & GTTGTAGTGGTATGAGGTTGGATTCATGGGCTTGAGATGGTTTTCATCAA           \\
8{[}136{]}6{[}152{]}\_MB   & GTTGTAGTGGTATGAGGTTGAACTAACGCATTGTGAATTACCTTGTGTACAG         \\
23{[}128{]}25{[}128{]}\_MB & GTTGTAGTGGTATGAGGTTGAAAAAAAGAGCCTCCGGCCAGAGCGCAGGCGC         \\
23{[}96{]}25{[}96{]}\_MB   & GTTGTAGTGGTATGAGGTTGGGTTGTGTAACGTGCCGGACTTGTGCTGGAGG         \\
7{[}160{]}8{[}168{]}\_MB   & GTTGTAGTGGTATGAGGTTGTTTAAGAAGAAAAATCTACGTTAATAAAACG          \\
7{[}192{]}8{[}200{]}\_MB   & GTTGTAGTGGTATGAGGTTGCACCGGAACAAAATCACCGGAACCGTCAGGAC         \\
7{[}225{]}8{[}232{]}\_MB   & GTTGTAGTGGTATGAGGTTGTCAGAACCCCTTATTAGCGTTTGCCATCTTT          \\
26{[}200{]}24{[}216{]}\_MB & GTTGTAGTGGTATGAGGTTGGCACTAAACATTTGAGGATTTACTACCATA           \\
7{[}256{]}9{[}255{]}\_MB   & GTTGTAGTGGTATGAGGTTGTCAGAGCCCGCGTTTTCATCGGCCATTTGG           \\
10{[}232{]}8{[}248{]}\_MB  & GTTGTAGTGGTATGAGGTTGTTAAAGGTCGTCACCGACTTGAGCATTTTCGG         \\
22{[}168{]}21{[}192{]}\_MB & GTTGTAGTGGTATGAGGTTGCAGTGCCAGTCACGACGTTGTAAAAATTAATT         \\
10{[}264{]}8{[}280{]}\_MB  & GTTGTAGTGGTATGAGGTTGAGGGAAGGGCCAGCAAAATCACCATAGCGTCA         \\
10{[}296{]}8{[}312{]}\_MB  & GTTGTAGTGGTATGAGGTTGGACAAAATAGCAAGGCCGGAAATAATCAGT           \\
21{[}64{]}23{[}63{]}\_MB   & GTTGTAGTGGTATGAGGTTGTTGGGAAGTGAGAGATAGACTTTAAACTTA           \\
22{[}296{]}20{[}312{]}\_MB & GTTGTAGTGGTATGAGGTTGAAGAAAATAACCTCCGGCTTAGGAACGCGA           \\
22{[}264{]}20{[}280{]}\_MB & GTTGTAGTGGTATGAGGTTGTCATTTGAATCAAAATCATAGGTCAGTTAATT         \\
22{[}232{]}20{[}248{]}\_MB & GTTGTAGTGGTATGAGGTTGTACATAAATAAGACGCTGAGAAGAGGTTTGAA         \\
9{[}64{]}11{[}63{]}\_MB    & GTTGTAGTGGTATGAGGTTGCATAACCCCCCCCTCAAATGCTTGACTATT           \\
7{[}288{]}9{[}288{]}\_MB   & GTTGTAGTGGTATGAGGTTGGCCGCCAGGAATCAAGTTTGCCTTGTAGCACC         \\
17{[}96{]}19{[}96{]}\_MB   & GTTGTAGTGGTATGAGGTTGAAAGGCTATCATATGTACCCCGGTAAATAATT    \\
\\
\caption{{\bf Names and sequences of strands used for the origami in Fig.~\ref{fig:si_scadnano}.} \label{table:sequences}}
\end{longtable}

\clearpage

\section{Kinetic Hook Effect}
Unlike the case for the usual high dose hook effect\cite{HighDoseHookBook2013}, where a single equilibrium binding process operates, the kinetic hook effect requires a second process to convert sensors from the closed, signal-on state, to the open signal-off state. We hypothesize that here this process is DNA strand displacement. Further we hypothesize that analyte binding to the lily pad is slower than intramolecular closure (where the analyte engages a second binder on the sensor). Thus our picture is that fast binding of an analyte molecule to one binder of the lily pad is followed by faster intermolecular closure. At high concentration this results in an out-of-equilibrium situation which can be relaxed by the slow strand displacement of a single DNA analyte within a closed lily pad, and which results in an open lily pad with two bound analytes. On this picture, the kinetic hook effect is likely intrinsic to the DNA analyte rather than our sensor.  For analytes that do not have displacement mechanisms it may not be observed; we have not observed this effect for two protein analytes (streptavidin and PDGF-BB) but we may not have examined high enough analyte concentrations. We note that for these DNA-sensing lily pads, the kinetic hook effect limits the upper limit of quantification, rather than signal saturation.

\section{Optimizing the Streptavidin Sensor}

\begin{table}[h]
{\Large \bf{A}} \\
\vspace*{-0.2in}
\begin{center}
\begin{tabular}{|c c|c|c|c|c|}
\hline
 \multirow{2}{*}{\hspace*{0.1in} $\delta L$ / MPC} & \multirow{2}{*}{$\searrow$} & \multicolumn{4}{c|}{Surface linker $L$} \\
  & & \multicolumn{1}{c}{14} &  \multicolumn{1}{c}{24} &  \multicolumn{1}{c}{34} &  \multicolumn{1}{c|}{44}\\ \hline
  \multirow{3}{*}{$l_{MB}$} & 20 & \cellcolor[HTML]{CBCEFB} 7.72 nm / 1.25 nA  & \cellcolor[HTML]{FFCCC9} 11.12 nm / 1.65 nA & \cellcolor[HTML]{EDEDED} 14.52 nm / 1.9 nA & \cellcolor[HTML]{EDEDED} 17.92 / 1.0 nA \\ 
   & 40 & \cellcolor[HTML]{96FFFB} 0.92 nm / 2.1 nA & \cellcolor[HTML]{FFFFC7} 4.32 nm / 2.75 nA & \cellcolor[HTML]{CBCEFB} 7.72 nm / 3.5 nA & \cellcolor[HTML]{FFCCC9} 11.12 nm / 1.1 nA \\ 
   & 60 & \cellcolor[HTML]{EDEDED} -5.88 nm / 0.8 nA & \cellcolor[HTML]{EDEDED} -2.48 nm / 1.5 nA & \cellcolor[HTML]{96FFFB} 0.92 nm / 2.6 nA & \cellcolor[HTML]{FFFFC7} 4.32 nm / 1.25 nA \\ 
    \hline
\end{tabular}
\end{center}

\vspace*{0.2in}
{\Large \bf{B}} \\
\vspace*{-0.2in}

\begin{center}
\begin{tabular}{|c c|c|c|c|c|}
\hline
\multirow{2}{*}{\hspace*{0.1in} $\delta L $ / MPC} & \multirow{2}{*}{$\searrow$} & \multicolumn{4}{c|}{Surface linker $L$} \\
  & & \multicolumn{1}{c}{14} &  \multicolumn{1}{c}{24} &  \multicolumn{1}{c}{34} &  \multicolumn{1}{c|}{44}\\ \hline
  \multirow{3}{*}{$l_{MB}$} & 20 & \cellcolor[HTML]{FFCE93} 7.72 nm / 1.25 nA  & \cellcolor[HTML]{FFCE93} 11.12 nm / 1.65 nA & \cellcolor[HTML]{DAE8FC} 14.52 nm / 1.9 nA & \cellcolor[HTML]{DAE8FC} 17.92 / 1.0 nA \\ 
   & 40 & \cellcolor[HTML]{9AFF99} 0.92 nm / 2.1 nA & \cellcolor[HTML]{9AFF99} 4.32 nm / 2.75 nA & \cellcolor[HTML]{9AFF99} 7.72 nm / 3.5 nA & \cellcolor[HTML]{FFCE93} 11.12 nm / 1.1 nA \\ 
   & 60 & \cellcolor[HTML]{DAE8FC} -5.88 nm / 0.8 nA & \cellcolor[HTML]{DAE8FC} -2.48 nm / 1.5 nA & \cellcolor[HTML]{FFCE93} 0.92 nm / 2.6 nA & \cellcolor[HTML]{9AFF99} 4.32 nm / 1.25 nA \\ 
    \hline
\end{tabular}
\caption{\vspace*{-0.0in} {\bf Effects of Analyte-binder Complex Size and Curtain length.\label{table:distances}} ({\bf A}) Each cell reports the difference, $\delta L$, between analyte-binder stack size and the MB curtain length where $\delta L = $~[$.34 \times (L+14)+ 5$]~-~$l_{MB}$~nm, and the maximum peak current (MPC) for an experimental condition having surface linker of size $L$ and MB curtain of length $l_{MB}$. Four pairs of distinct experimental conditions with the same $\delta L$ (0.92~nm, blue; 4.32~nm, yellow; 7.72~nm, purple; 11.12~nm, red) but different MPC falsify the hypothesis that $\delta L$ is the only determinant of MPC. For these pairs in which $\delta L$ is maintained constant by increasing $L$ and $l_{MB}$ by 20~bp (6.8~nm) in lockstep, sometimes the MPC goes up (purple, blue) and sometimes the MPC goes down (red, yellow). The relatively large MPC of 1.9~nA for a relatively large $\delta L$ of 14.52~nm (for $L=34$ and $l_{MB}=20$) suggests that a linker length above a threshold length is required to prevent steric interaction of the origami shape with the surface. We suggest that this threshold is determined by some characteristic static deformation or dynamic fluctuation of the origami shape at the 20~nm length scale. ({\bf B}) $\delta L$ and MPC are reported as before, but cell coloring is different. Within each surface linker column, green indicates highest MPC, orange the second highest MPC, and light blue the lowest MPC. As $L$ increases, the rank of the $l_{MB} = 60$ curtain increases from lowest to highest MPC. These data suggest that as $L$ increases and the origami lifts farther from the surface, steric interference from the MB decreases, and in the limit that steric interference from the origami ceases, the 60~bp MB curtain works best, because it reaches closer to the surface. }
\end{center}
\end{table}